\documentclass[aps,prd,10pt,floatfix,twocolumn,
               superscriptaddress,preprintnumbers,
               shownopacs,showkeys,nofootinbib,
               notoccite,notitlepage]{revtex4-1}

\usepackage[utf8]{inputenc}
\usepackage[T1]{fontenc}
\usepackage{microtype}

\usepackage{amsmath,amssymb,bm}
\usepackage{siunitx}

\usepackage{graphicx}
\usepackage{dcolumn}
\usepackage{float}
\usepackage{footmisc}

\usepackage{xcolor}

\usepackage[colorlinks=true,linkcolor=blue,
           citecolor=blue,urlcolor=blue]{hyperref}
\usepackage{aas_macros}

\preprint{}

\begin{document}
\rightline{FERMILAB-PUB-26-0309-T}
\title{Formation and Redshift Evolution of Dark Matter Spikes}

\author{Gonzalo Herrera}
\email{gonzaloh@mit.edu}
\thanks{ORCID: \href{https://orcid.org/0000-0001-9250-8597}{0000-0001-9250-8597}}
\affiliation{Department of Physics and Kavli Institute for Astrophysics and Space Research, MIT, Cambridge, MA 02139, USA}
\affiliation{Harvard University, Department of Physics and Laboratory for Particle Physics and Cosmology, Cambridge, MA 02138, USA}
\author{Abdelaziz Hussein}
\email{abdelh@mit.edu}
\affiliation{Department of Physics and Kavli Institute for Astrophysics and Space Research, MIT, Cambridge, MA 02139, USA}
\author{Lina Necib}
\affiliation{Department of Physics and Kavli Institute for Astrophysics and Space Research, MIT, Cambridge, MA 02139, USA}
\author{Elliot Y. Davies}
\affiliation{Department of Physics and Kavli Institute for Astrophysics and Space Research, MIT, Cambridge, MA 02139, USA}
\author{Xuejian Shen}
\affiliation{Department of Physics and Kavli Institute for Astrophysics and Space Research, MIT, Cambridge, MA 02139, USA}

\begin{abstract}
Dark matter density spikes forming around adiabatically growing black holes can dramatically enhance indirect and direct detection signals. 
Canonical predictions, however, assume a zero-mass seed in a purely dark matter environment and do not track the long-term dynamical impact of surrounding stars. 
We present a semi-analytic framework that first generalizes adiabatic spike formation to include finite seed masses, stellar cusps, and non-circular orbits, and then studies the subsequent cosmic evolution by solving coupled Fokker--Planck equations for the dark matter and stellar phase-space distributions, with a heating rate modulated by the cosmic star formation rate. 
Starting conservatively from canonical Gondolo--Silk spikes and marginalizing over astrophysical uncertainties, we find that stellar gravitational heating drives the inner slope towards $\gamma_\chi \simeq 1.5$ within a few Gyrs (\textit{e.g} by $z \lesssim 2$ for spikes formed at $z\simeq 10$), yielding overdensities two to four orders of magnitude below canonical expectations but still well above an NFW-like cusp. 
We provide redshift-dependent benchmarks for the column density and $J$-factor relevant to scattering, decay and annihilation signatures. Any robust interpretation of indirect dark matter signals from galactic nuclei must account for this evolution.
\end{abstract}

\maketitle

\section{Introduction}

The existence of dark matter (DM) is firmly established through its gravitational effects across a wide range of physical scales, from the cosmic microwave background \cite{Hinshaw_2009,2013ApJS..208...20B} and galaxy clustering to the rotation curves of galaxies \cite{1980ApJ...238..471R,1937ApJ....86..217Z} (see \cite{Bertone_2018,cirelli2025darkmatter} for reviews). On sub-parsec scales, however, the distribution of DM in the vicinity of supermassive black holes (SMBHs) remains poorly constrained. Measurements of stellar orbits near Sagittarius~$A^{\star}$ show no evidence for an enclosed DM mass comparable to the black hole mass, placing upper limits on the DM density profile at sub-pc scales \cite{Lacroix:2018zmg,Shen:2023kkm}. Mild hints of DM overdensities around black holes have emerged from reverberation mapping of Active Galactic Nuclei \cite{Sharma:2025ynw} and from orbital-period measurements of black hole binaries \cite{Alachkar:2022qdt,chan2024robustevidenceshowingdark, Deb:2025raq, Scarcella:2025dsh}, though these remain limited by uncertainties in the intrinsic black hole masses.

On the theoretical side, Gondolo and Silk~\cite{Gondolo:1999ef} showed that adiabatic growth of a SMBH at the center of a DM halo produces a steep density cusp, usually dubbed a ``DM spike,'' with inner slope $\gamma_{\rm sp} = (9-2\gamma)/(4-\gamma)$ for an initial halo profile $\rho \propto r^{-\gamma}$. This result, built on earlier work on baryon contraction in the vicinity of central black holes \cite{1972GReGr...3...63P,1980ApJ...242.1232Y,kohn_kulrsrud,Quinlan:1994ed}, assumes that the SMBH grows slowly compared to the local dynamical timescale, a condition well satisfied for Salpeter growth timescales of $t_{\rm sap}\sim 5\times 10^7$\,yr compared to typical orbital timescales of $t_{\rm dyn}\sim 10^4$--$10^5$\,yr \cite{Sigurdsson:2003wu, Herrera:2023nww, 1964ApJ...140..796S}. The presence of such spikes would significantly enhance indirect and direct DM signals: self-annihilation fluxes \cite{Gondolo:1999ef, Lacroix:2015lxa,Shelton:2015aqa,Shapiro:2016ypb, Chattopadhyay:2026kbm,Phoroutan-Mehr:2024cwd,Kivokurtseva:2024prw, BetancourtKamenetskaia:2025ivl}, cosmic-ray cooling \cite{Herrera:2023nww, DeMarchi:2024riu,Gustafson:2024aom,Mishra:2025juk, Kantzas:2025huu, Hussein:2025llu, Meighen-Berger:2025hrq}, cosmic-ray boosted DM \cite{Wang:2021jic,Gustafson:2025dff, DeMarchi:2025uoo}, gamma-ray \cite{Ferrer:2022kei, Herrera:2025gpm} and neutrino emissions \cite{Ferrer:2022kei,Cline:2022qld, Fujiwara:2023lsv, Pompa:2025lbf, Zapata:2025huq, Fujiwara:2024qos, Tseng:2024akh, Mondol:2025uuw, Bertolez-Martinez:2025trs}, as well as imprints on gravitational waveforms from compact binary inspirals \cite{Alonso-Alvarez:2024gdz, Bertone:2024rxe, Sharpe:2026nqq}, or changes in the orbital decay or stellar motion induced by non-gravitational interactions of DM with stars \cite{Acevedo:2025rqu, Gustafson:2025ypo}.

Despite their phenomenological importance, existing predictions for DM spikes rely on several simplifying assumptions that limit their applicability to realistic galactic nuclei. The canonical Gondolo--Silk prescription assumes a zero-mass seed black hole, a purely DM initial potential, and circular particle orbits. The subsequent dynamical evolution of spikes due to stellar gravitational heating was studied for Sagittarius~$A^{\star}$ in \cite{Merritt:2003qk,Merritt:2006mt,Merritt:2002vj, Vasiliev:2008uz}, which showed that gravitational encounters with stars can significantly soften the inner cusp. However, these works assumed a static stellar profile with a time-independent heating rate, did not account for the effect of a surrounding stellar population or a finite seed mass on the formation of the spike, and did not track the redshift evolution of the stellar component self-consistently, further accounting for varied initial conditions of the system. To date, no study has combined a realistic spike formation prescription with a redshift-dependent treatment of the subsequent two-component dynamical evolution, nor has discussed the impact of the spike redshift evolution for multiple initial conditions.
\begin{figure*}[t!]
  \centering
\includegraphics[width=0.49\linewidth]{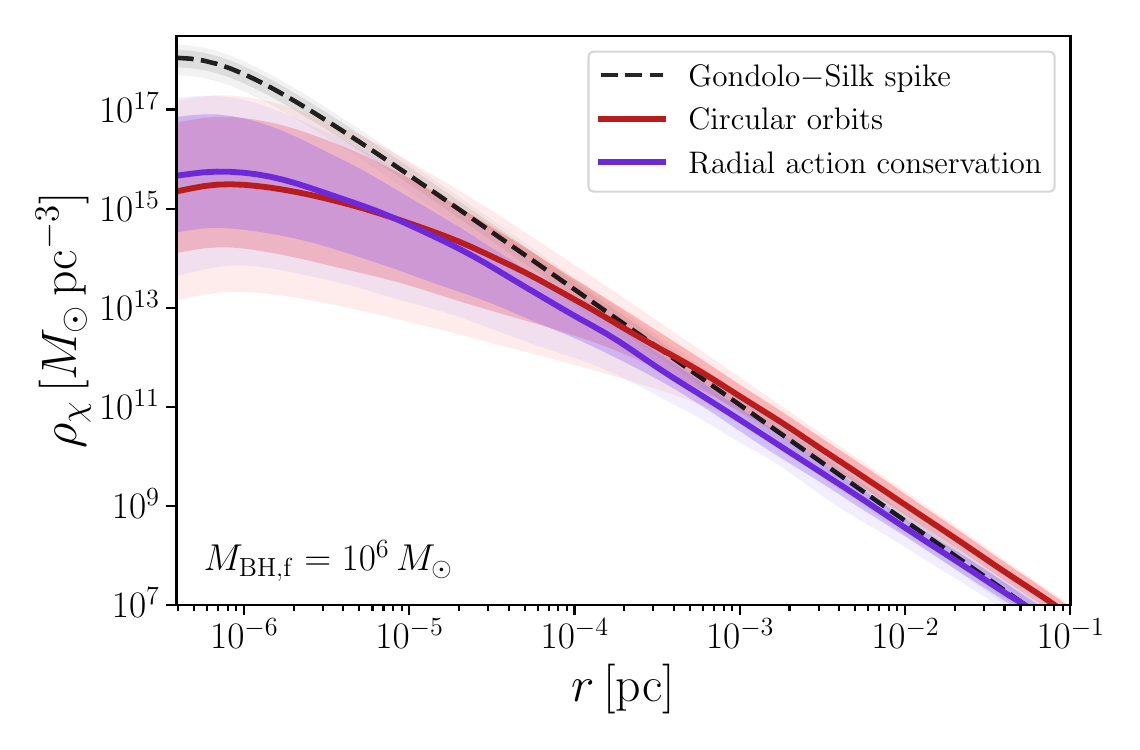}
\includegraphics[width=0.49\linewidth]{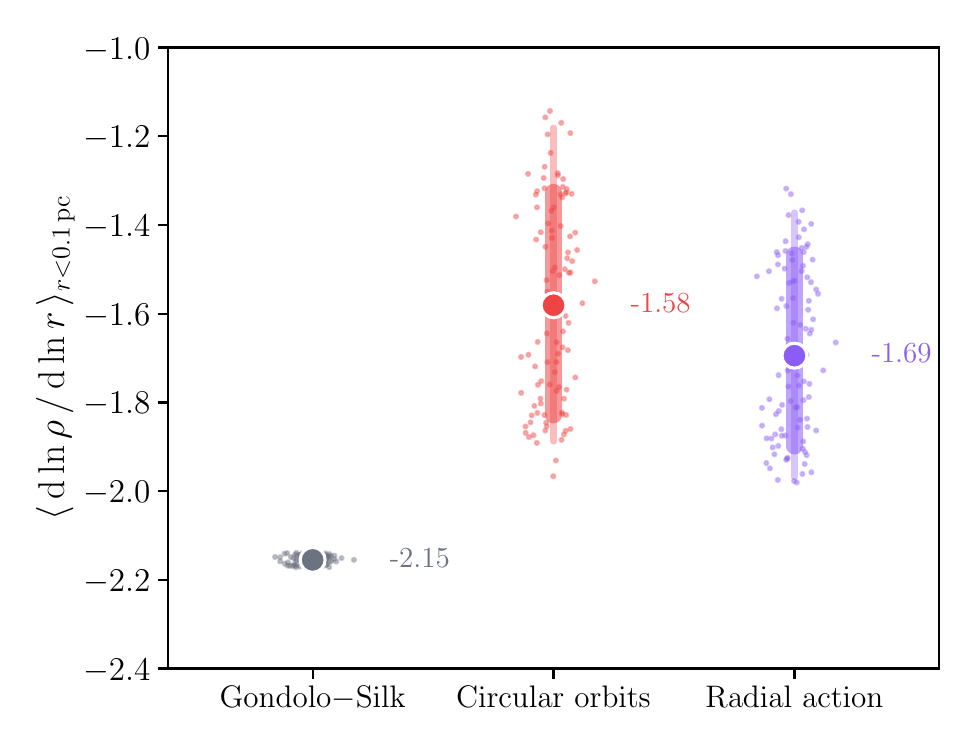}

  \caption{\textit{Left panel:} DM density profiles $\rho_\chi(r)$ after adiabatic contraction driven by the growth of a black hole to a final mass $M_{\rm BH,f}=10^{6}\,M_\odot$, comparing three spike formation prescriptions. Each is obtained from a four-dimensional Monte Carlo marginalization over log-uniform priors on (i) the initial dark-matter cusp slope $\gamma_{\chi,0}\in[0.85,1.15]$, (ii) the initial stellar cusp slope $\gamma_{\star,0}\in[1.2,1.8]$, (iii) the enclosed stellar-to--dark-matter mass ratio $\eta\equiv M_\star(<1\,\mathrm{pc})/M_\chi(<1\,\mathrm{pc})\in[0.3,3]$, and (iv) the seed black-hole mass $M_{\rm BH,i}\in[10^{2.5},10^{4.5}]\,M_\odot$. Gray bands show the Gondolo--Silk prescription (black dashed median), red bands the circular-orbit adiabatic contraction including a finite seed mass and stellar cusp (dark red median), and purple bands the full radial-action conservation method (purple median). Dark (light) bands indicate the $68\%$ ($95\%$) credible intervals. \textit{Right panel:} Averaged logarithmic slope $\langle\mathrm{d}\ln\rho_\chi/\mathrm{d}\ln r\rangle$ within $r<0.1\,\mathrm{pc}$ for the same three prescriptions. Thick bars show the $68\%$ interval, thin bars the $95\%$ interval, and circles denote the median. The canonical Gondolo--Silk prediction yields the steepest slopes, radial-action conservation yields softer profiles once finite seeds and stars are included, and circular-orbit adiabatic contraction produces the shallowest inner profiles.}

  \label{fig:cosmic_median_formation}
\end{figure*}

In this work we address these limitations in two steps. First, we generalize the adiabatic contraction prescription for spike formation to include a finite black hole seed mass, a surrounding stellar cusp with arbitrary normalization and slope, and non-circular DM orbits treated via full radial-action conservation. We show that the resulting spikes are sizably softer than the canonical Gondolo--Silk prediction once stars and finite seeds are accounted for. Second, we study the subsequent cosmic evolution of these overdensities by solving coupled, orbit-averaged Fokker--Planck equations for the DM and stellar phase-space distributions, with a heating rate that tracks the cosmic star formation history. For the evolution, we adopt canonical Gondolo--Silk spikes as initial conditions, which represent the steepest and most optimistic starting point; the relaxation timescales we derive are therefore conservative. We marginalize over the uncertain initial conditions, including the seed mass, stellar and DM profile slopes, and enclosed mass ratios, to provide robust predictions for the inner DM slope as a function of redshift.

We do not attempt to model gas dynamics, black hole mergers, or structure formation in a cosmological context; our goal is to isolate the dominant gravitational processes controlling the survival of DM overdensities in dense stellar environments.

The remainder of this paper is organized as follows. In Sec.~\ref{sec:formation}, we present our generalized spike formation framework, examining the dependence of the resulting DM overdensities on the seed mass, stellar profile, enclosed mass ratios, and orbital assumptions. In Sec.~\ref{sec:evolution_I}, we introduce the Fokker--Planck formalism for the subsequent dynamical evolution and derive the analytical steady-state solutions. In Sec.~\ref{sec:evolution_II}, we simulate the redshift evolution of DM spikes in a fixed stellar bath, both with constant normalization and with a star-formation-rate-dependent heating rate. In Sec.~\ref{sec:evolution_III}, we present results from the fully coupled coevolution of the DM and stellar distributions. In Sec.~\ref{sec:sfr_impact}, we assess the impact of bursty star-formation histories. In Sec.~\ref{sec:formation_z}, we examine the sensitivity of the evolved profiles to the assumed spike formation redshift. In Sec.~\ref{sec:implications}, we discuss the implications for indirect DM searches. Finally, in Sec.~\ref{sec:conclusions}, we summarize our conclusions and outline future directions.
\section{Formation of DM overdensities within a star cluster}\label{sec:formation}

In this section, we focus exclusively on the formation stage of DM overdensities during black hole growth. We do not yet include dynamical heating or relaxation effects from stars, which are treated in Secs.~\ref{sec:evolution_I}--\ref{sec:evolution_III}. We first review the canonical Gondolo--Silk framework (Sec.~\ref{sec:canonical_spike}), then generalize it to include finite seed masses and stellar potentials assuming circular orbits (Sec.~\ref{sec:generalized_circular}), and finally present a full treatment based on radial-action conservation for non-circular orbits (Sec.~\ref{sec:radial_action}).

\subsection{Canonical spike formation}\label{sec:canonical_spike}

The traditional framework for the formation of dark-matter (DM) overdensities around a supermassive black hole (SMBH) is the adiabatic growth model from~\cite{Quinlan:1994ed, Gondolo:1999ef}. In this scenario, the SMBH grows slowly relative to the orbital period of DM particles, such that the adiabatic invariant for circular orbits,
\begin{equation}\label{eq:adiabatic_invariant}
    r\,M(<r)=\mathrm{const},
\end{equation}
is conserved during the growth. Here $r$ denotes the orbital radius of a DM particle, and $M(<r)$ the total enclosed mass within that radius. Assuming an initial cuspy DM profile,
\begin{equation}
    \rho_\chi(r)=\rho_0\Bigl(\frac{r}{r_0}\Bigr)^{-\gamma},
\end{equation}
where $\rho_0$ is a reference density and $r_0$ is the halo scale radius, with $\gamma=1$ for an NFW-like halo~\cite{Navarro:1995iw}, conservation of $r\,M(<r)$ maps each initial radius $r_i$ to a final radius
\begin{equation}
    r_f = r_i \frac{M_{\chi,i}(r_i)}{M_{\rm BH}},
\end{equation}
where $M_{\chi,i}(r)$ is the initial enclosed DM mass within radius $r$ and $M_{\rm BH}$ is the final SMBH mass. This mapping leads to a steepening of the DM density profile to an inner slope $\gamma_{\rm sp}=\frac{9-2\gamma}{4-\gamma}$ (yielding $\gamma_{\rm sp}=7/3$ for $\gamma=1$).

The above argument fixes the asymptotic inner slope of the overdensity under adiabatic black-hole growth, but does not by itself determine the overall normalization of the spike or the radial extent over which it applies. In Ref.~\cite{Gondolo:1999ef}, these quantities are obtained by explicitly mapping the initial dark-matter phase-space distribution into the final black-hole-dominated potential, enforcing conservation of phase-space density and angular momentum during the adiabatic growth. This treatment yields both the characteristic spike radius $R_{\rm sp}$, defined by the transition between the contracted inner profile and the original halo, and the normalization $\rho_R$ obtained by matching the spike to the initial density profile at $R_{\rm sp}$. Concretely, the spike density profile reads (see Appendix~\ref{app:GS_derivation} for further details)
\begin{equation}
    \rho_\chi(r)=
    \rho_R\, g_\gamma(r)\,
    \left(\frac{R_{\mathrm{sp}}}{r}\right)^{\gamma_{\mathrm{sp}}},
    \label{eq:GSprofile}
\end{equation}
where $\rho_R = \rho_0(R_{\rm sp}/r_0)^{-\gamma}$ is a normalization constant chosen so that the spike matches the original profile at $R_{\rm sp}$, and the spike extends out to
\begin{equation}\label{eq:spike_radius}
    R_{\mathrm{sp}}
    = \alpha_\gamma\, r_0\left(\frac{M_{\rm BH}}{\rho_0 r_0^3}\right)^{1/(3-\gamma)},
\end{equation}
with $\alpha_\gamma$ taking numerical values in the range $\alpha_{\gamma} \simeq0.007$--$0.018$ for $\gamma=0.05$--$2$~\cite{Gondolo:1999ef}. The factor $g_\gamma(r)\simeq \left(1-{2R_S}/{r}\right)^{3}$ accounts for relativistic capture, smoothly vanishing the density at $r=2R_S$, where $R_S=2GM_{\rm BH}/c^2$ is the Schwarzschild radius \cite{Sadeghian:2013laa}.

This model relies on several simplifying assumptions: adiabatic BH growth, a pure DM initial potential with no baryons, a zero-mass seed, and a final potential dominated by the BH. We revisit each of these assumptions below and assess their impact on the final DM profiles. The Gondolo--Silk prescription serves here primarily as a benchmark against which more realistic formation scenarios will be compared.

\subsection{Generalized formation with finite seed mass and stellar cusp}\label{sec:generalized_circular}

We now extend the canonical spike formation to incorporate a realistic stellar nucleus and a finite BH seed mass. The initial mass profile entering the adiabatic invariant is
\begin{equation}
    M_{\rm tot,i}(r) = M_{\rm BH,i}+M_{\chi,i}(r)+M_{\star,i}(r),
\end{equation}
with
\begin{align}
\rho_\chi(r) &= \rho_{\chi,0}\left(\frac{r}{r_0}\right)^{-\gamma_\chi}\left(1-\dfrac{2R_S}{r}\right)^{3},\\[3pt]
\rho_\star(r)&= \rho_{\star,0}\left(\frac{r}{r_0}\right)^{-\gamma_\star}\left(1-\dfrac{2R_S}{r}\right)^{3}.
\end{align}

Assuming circular orbits, the generalized invariant becomes
\begin{equation}
   r_i M_{\rm tot,i}(r_i)=r_f\,M_{\rm BH,f},
\end{equation}
yielding the mapping
\begin{equation}
   r_f=r_i\,\frac{M_{\rm tot,i}(r_i)}{M_{\rm BH,f}}.
\end{equation}
This reduces exactly to the canonical relation when stars are absent and $M_{\rm BH,i}\to0$. We consider that each shell $\mathrm{d}M_\chi$ at radius $r_i$ is mapped to a new radius $r_f$. The final density profile is then reconstructed by distributing the shell masses into logarithmic bins. In particular, we solve for
\begin{equation}
\rho_{\chi,f}\left(r_f\right)=\rho_{\chi, i}\left(r_i\left(r_f\right)\right)\left(\frac{r_i}{r_f}\right)^2\left|\frac{d r_i}{d r_f}\right|.
\end{equation}

The final DM density profile is subject to various uncertainties such as the initial black hole seed mass $M_{\rm BH,i}$, the stellar density normalization $\rho_{\star,0}$, steepness $\gamma_{\star}$, and the final DM density profile normalization $\rho_{0,\chi}$ and steepness $\gamma_{\chi}$. We consider seed masses in the ranges $
10^{2.5}\,M_\odot \;\leq\; M_{\rm BH,i} \;\leq\; 10^{4.5}\,M_\odot $, which spans the theoretically motivated spectrum of black-hole formation channels: remnants of Population-III stars ($\sim 10^{2}\,M_\odot$) \cite{Madau:2001sc}, runaway stellar collisions in dense nuclear clusters ($10^{3}$-$10^{4}\,M_\odot$) \cite{PortegiesZwart:2002iks}, and direct-collapse black holes ($\sim 10^{4}$-$10^{6}\,M_\odot$) \cite{Begelman:2006db}. We allow the stellar index to vary within $\gamma_{\star}=1.2-1.8$ \cite{1976ApJ...209..214B, Genzel_2010, Genzel:2003cn, Preto:2004kd}, the initial DM profile index within $\gamma_{\chi}=0.85-1.15$ \cite{Navarro:1995iw, Navarro:1996gj}, and the enclosed stellar-to--dark-matter mass ratio within $1\,\mathrm{pc}$, $\eta\equiv M_\star(<1\,\mathrm{pc})/M_\chi(<1\,\mathrm{pc})\in[0.3,3]$, consistent with resolved stellar-dynamics measurements in the Galactic Center \cite{Genzel_2010, 2003ApJ...596.1015S} and nuclear star cluster mass estimates in nearby galaxies \cite{2020A&ARv..28....4N}, which indicate enclosed stellar-to-dark-matter mass ratios of order unity within the influence radius. The BH grows to a final mass of $M_{\rm BH}=10^{6} M_{\odot}$. To sample over these uncertain ranges, we perform a Monte Carlo marginalization with log-uniform priors over the aforementioned ranges for each of the three formation prescriptions: the canonical Gondolo--Silk formula, the circular-orbit adiabatic contraction (this section), and the radial-action conservation method (Sec.~\ref{sec:radial_action}).

The results are shown in Fig.~\ref{fig:cosmic_median_formation}. The left panel displays the marginalized density profiles for all three prescriptions. The Gondolo--Silk bands (gray) lie at the highest densities at small radii, reflecting the canonical spike slope $\gamma_{\rm sp}\simeq 7/3$ in the absence of stars and with a zero seed mass. The circular-orbit adiabatic contraction (red bands) produces systematically lower densities at radii $r \lesssim 10^{-4}$\,pc due to the finite seed mass and stellar cusp, with the cosmic median differing from the Gondolo--Silk result by up to $\sim 2$ orders of magnitude. The radial-action conservation method (purple bands) yields intermediate densities in the innermost region compared to the canonical spike scenario and the circular orbits scenario. Despite these reductions, the resulting DM profiles remain very dense within the region of gravitational influence of the black hole, well above an extrapolated NFW cusp. The right panel summarizes these differences through the averaged inner slope within $r<0.1\,\mathrm{pc}$, confirming that the inclusion of a finite seed mass, stellar cusp, and non-circular orbits systematically softens the inner profile relative to the canonical prediction. In Appendix~\ref{app:formation}, we isolate the impact of varying each of the uncertain parameters and discuss their effect on the spike formation.

\subsection{Non-circular orbits: radial-action conservation}\label{sec:radial_action}

The circular-orbit prescription from Eq.~\ref{eq:adiabatic_invariant} is only approximate and may not accurately capture the dynamics in the vicinity of the SMBH, where the potential receives contributions from the black hole, stars, and DM. A proper adiabatic treatment should instead rely on the radial action \cite{Bertone:2024wbn}
\begin{equation}
J_r(E,L)=\frac{1}{\pi}\int_{r_{\rm p}}^{r_{\rm a}}\mathrm{d}r\,
\sqrt{2\,[E-\Phi(r)]-\frac{L^2}{r^2}},
\label{eq:Jr_def}
\end{equation}
which is an exact adiabatic invariant for collisionless systems in spherical symmetry. 
Here \(E\) and \(L\) are the orbital energy and angular momentum, and 
\(r_{\rm p}\) and \(r_{\rm a}\) denote the pericenter and apocenter, defined by the turning points of the radial motion. 
During slow SMBH growth the gravitational potential changes from its initial form \(\Phi_i(r)\) to its final form \(\Phi_f(r)\), 
while the angular momentum remains conserved. Each orbit then obeys
\begin{equation}
J_r^{\rm (i)}(E_i,L)=J_r^{\rm (f)}(E_f,L).
\label{eq:Jr_matching}
\end{equation}
In the SMBH-dominated regime the final potential approaches a Keplerian form,
\(\Phi_f(r)\simeq-G M_{\rm BH}/r\), for which the radial action of bound orbits (\(E_f<0\)) can be written in closed form,
\begin{equation}
J_r^{\rm (f)}(E_f,L)=\frac{G M_{\rm BH}}{\sqrt{-2E_f}}-L.
\label{eq:Jr_kepler}
\end{equation}
In practice, however, the final potential retains contributions from the DM and stellar distributions, so Eq.~\ref{eq:Jr_matching} must be solved numerically in general.

The initial composite potential \(\Phi_i(r)\) is constructed self-consistently from the extended mass distribution (seed SMBH, DM and stars), truncated at an outer radius \(r_{\max}=1~\mathrm{pc}\) and normalized such that \(\Phi_i(r_{\max})\) matches the enclosed-mass boundary condition. The isotropic velocity dispersion in this potential is obtained from the Jeans equation,
\begin{equation}
\rho(r)\,\sigma^2(r) = \int_r^{r_{\max}} \rho(r')\,\frac{G\,M(<r')}{r'^{\,2}}\,dr',
\label{eq:jeans_sigma}
\end{equation}
which provides the equilibrium velocity scale at each radius.

The final density profile is obtained by mapping initial orbits forward through action conservation via a Monte Carlo procedure \cite{Metropolis01091949}. We sample DM phase-space points \((r,\,\vec{v})\) in the initial composite potential. Positions are drawn uniformly in \(\log r\) and weighted by the local density. Velocities are drawn from an isotropic Gaussian with dispersion \(\sigma(r)\) obtained from Eq.~\ref{eq:jeans_sigma}, with particles exceeding the local escape velocity \(v_{\rm esc}(r)=\sqrt{-2\,\Phi_i(r)}\) redrawn until all samples are bound. From each sampled position and velocity we obtain the initial orbital energy \(E_i = \Phi_i(r) + \tfrac{1}{2}v^2\) and angular momentum \(L = r\,v_\perp\), with \(v_\perp^2 = v^2 - v_r^2\), and evaluate the radial action \(J_r^{(\rm i)}(E_i,L)\) numerically via Eq.~\ref{eq:Jr_def}. The final energy \(E_f\) is then determined by the matching condition \(J_r^{(\rm f)}(E_f,L)=J_r^{(\rm i)}\) at fixed \(L\), which we solve by bisection in the final potential. Once \(E_f\) is known, the contribution of each particle to the final density is distributed along its orbit using time averaging, weighting each radial bin by \(\Delta t \propto \Delta r / v_r\), where \(v_r(r)\) is the radial velocity in the final potential. The final density profile \(\rho_f(r)\) is then obtained by summing the weighted contributions over all particles.

The impact of non-circular orbits is shown for individual parameter choices in Appendix~\ref{app:formation}, Fig.~\ref{fig:full_radial_action}, and in the marginalized results of Fig.~\ref{fig:cosmic_median_formation}. Radial-action conservation yields slightly steeper inner slopes than the circular-orbit prescription, reflecting the richer orbital structure encoded in $J_r(E,L)$ compared to the single invariant $rM(r)$.

In the remainder of this paper, we adopt canonical Gondolo--Silk spikes as initial conditions for the subsequent dynamical evolution. As discussed above, this represents the steepest plausible starting point; the inclusion of finite seeds, stars, and non-circular orbits would only reduce the initial overdensity further.

\section{Cosmic Evolution of DM overdensities in a star cluster: The steady-state solutions}\label{sec:evolution_I}

In this and following sections we focus on the evolution of the DM overdensities after being formed. In the previous section we saw that for several plausible scenarios for the initial black hole mass seed and DM and stellar initial density profiles, steep overdensities can be formed, in some scenarios even resembling the Gondolo and Silk expectation from Ref. \cite{Gondolo:1999ef}, see Appendix \ref{app:formation}.

In order to introduce our procedure to model DM-stellar encounters, we begin with a pedagogical introduction to simulations of black holes embedded in a star cluster as first done by Bahcall and Wolf \cite{1976ApJ...209..214B, Frank:1976uy, 1983ApJ...264...24M}, and the simulations of black holes embedded in a combined DM and star cluster, as done independently by Gnedin and Primack \cite{Gnedin:2003rj}, and Merritt \cite{Merritt:2003qk}. Inside the BH region of gravitational influence, the potential can be approximated as Keplerian,
$\Phi(r)=-G M_{\rm BH}/r$. The isotropic, orbit-averaged density of states
and its cumulative are \cite{2008gady.book.....B}
\begin{equation}
\begin{aligned}
p(E) &= \frac{\pi}{2\sqrt{2}}\,(G M_{\rm BH})^{3}\,E^{-5/2}, \\
q(E) &= \int_0^{\infty} p(E')\, dE'
     = \frac{\pi}{3\sqrt{2}}\,(G M_{\rm BH})^{3}\,E^{-3/2}.
\end{aligned}
\label{eq:pq}
\end{equation}
with $E>0$ the binding energy. The isotropic Fokker–Planck (FP) equation in energy space is
\begin{equation}
\begin{aligned}
4\pi^{2} p(E)\,\frac{\partial f}{\partial t}
&= -\frac{\partial F_E}{\partial E}, \\
F_E(E,t)
&= -\,D_{EE}(E,t)\,\frac{\partial f}{\partial E}
   + D_E(E,t)\,f .
\end{aligned}
\label{eq:FP_master}
\end{equation}
where $F_E$ is the energy flux, $D_{EE}$ encodes diffusion, accounting for random energy kicks, and $D_E$ the drift accounting for systematic energy exchange. For a single stellar mass $M_\star$, gravitational encounters among stars give the
orbit–averaged coefficients \cite{1976ApJ...209..214B,2008gady.book.....B}
\begin{align}
D_{EE}^\star(E)
&= 64\pi^{4} G^{2} m_\star^{2}\ln\Lambda
\Bigg[
q(E)\!\int_{0}^{E} f_\star(E')\,dE'
\notag \\
&\quad
+ \int_{E}^{\infty} q(E')\,f_\star(E')\,dE'
\Bigg],
\label{eq:DEE_star}\\
D_{E}^\star(E)
&= \,64\pi^{4} G^{2} m_\star^{2}\ln\Lambda
\int_{0}^{E} p(E')\,f_\star(E')\,dE'.
\label{eq:DE_star}
\end{align}
with $\mathrm{ln}\Lambda \simeq 15$ denoting the Coulomb logarithm \cite{2008gady.book.....B}. A steady-state has $F_E=0$, and solving for a power law $f_\star\propto E^{k_\star}$ in a Kepler potential yields the Bahcall–Wolf solution $k_\star=\tfrac{1}{4}$ and therefore (see Appendix \ref{app:steady_state_solution} for a derivation)
\begin{equation}
f_\star(E)\propto E^{1/4}, \qquad \rho_\star(r)\propto r^{-(k_\star+3/2)}=r^{-7/4}.
\label{eq:BW}
\end{equation}

Collisionless DM interacts only gravitationally with the surrounding stars. In the limit where the DM particle mass is much smaller than the stellar mass, 
$m_\chi \ll m_\star$, the first-order (drift) term in the orbit-averaged Fokker-Planck equation 
for DM is parametrically suppressed compared to the second-order diffusion term as $\propto m_{\chi}/m_{\star}$, and can be safely neglected, $D_E^\chi \simeq 0$ (see Appendix~\ref{sec:diffusion_coefficients} for a derivation of the diffusion coefficients).

The diffusion term is the same operator driven by
the stellar bath, $D_{EE}^\chi \equiv D_{EE}^\star$. The DM steady state ($F_E^\chi=0$) thus implies (see Appendix \ref{app:steady_state_solution} for a derivation) \cite{Gnedin:2003rj,Merritt:2003qk}.
\begin{equation}
f_\chi(E)=\mathrm{const.}
\quad\Longrightarrow\quad \rho_\chi(r)\propto r^{-3/2}.
\label{eq:DM_32}
\end{equation}

\begin{figure*}[t!]
  \centering
\includegraphics[width=0.49\linewidth]{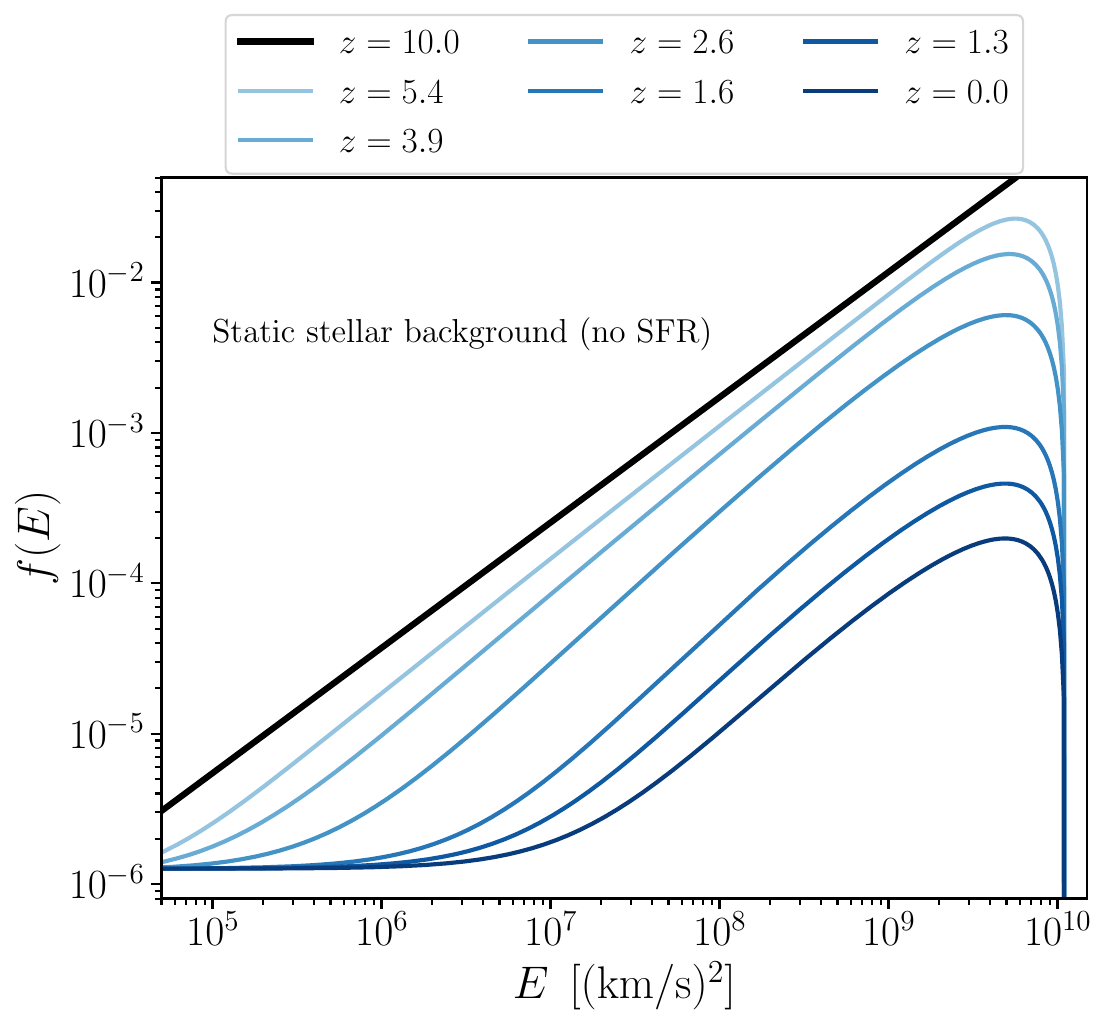}  \includegraphics[width=0.49\linewidth]{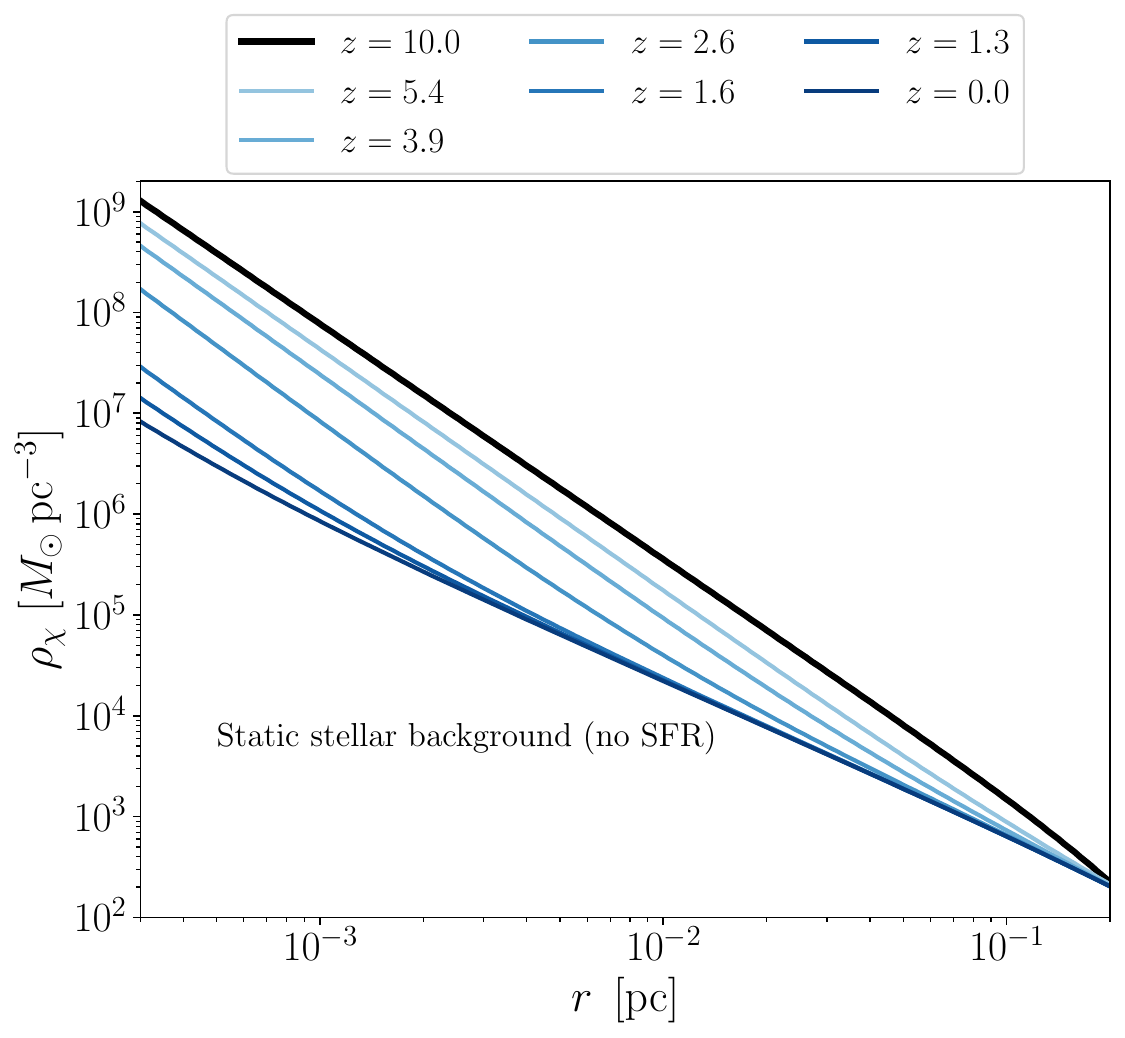}
\includegraphics[width=0.49\linewidth]{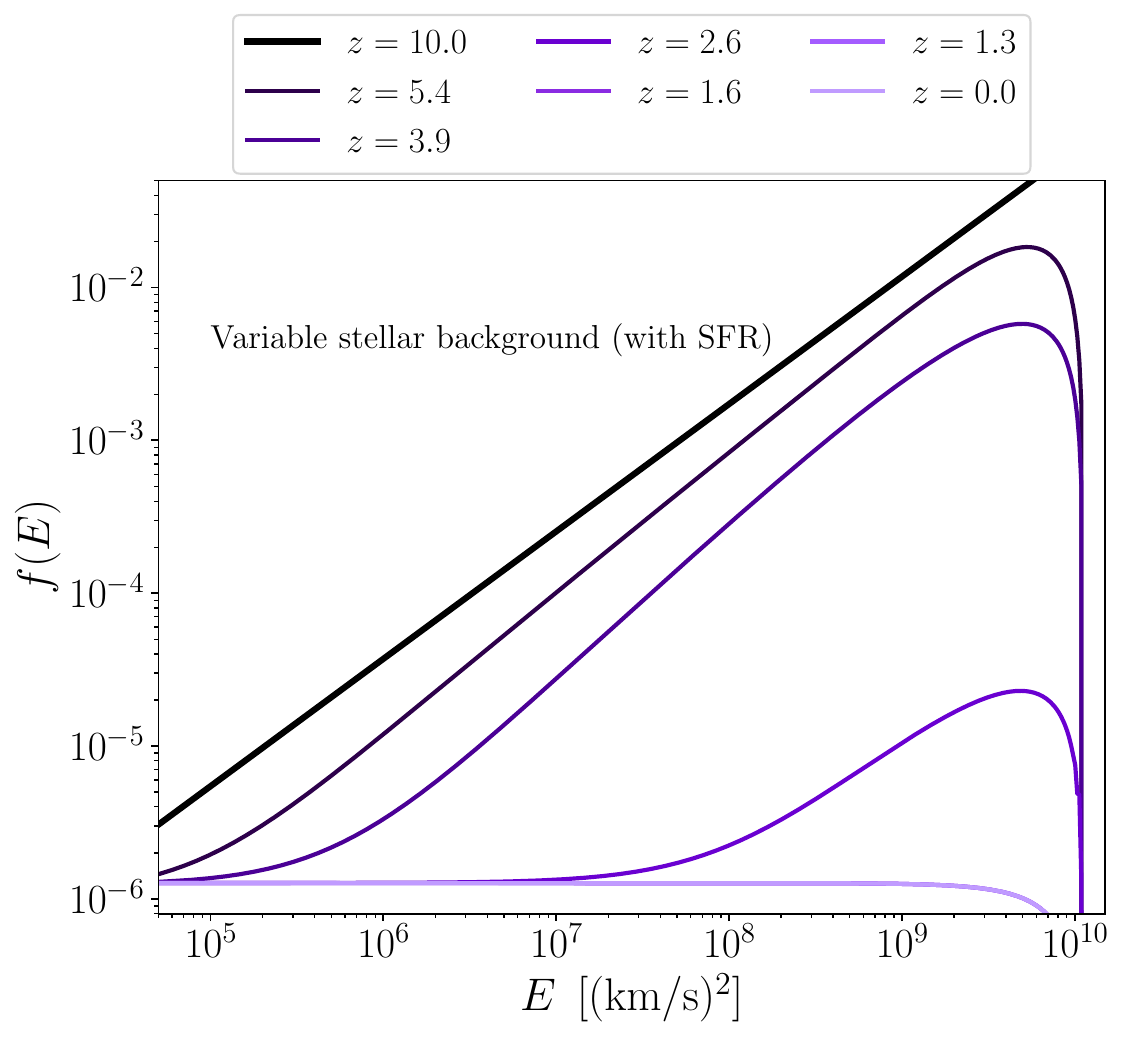}  \includegraphics[width=0.49\linewidth]{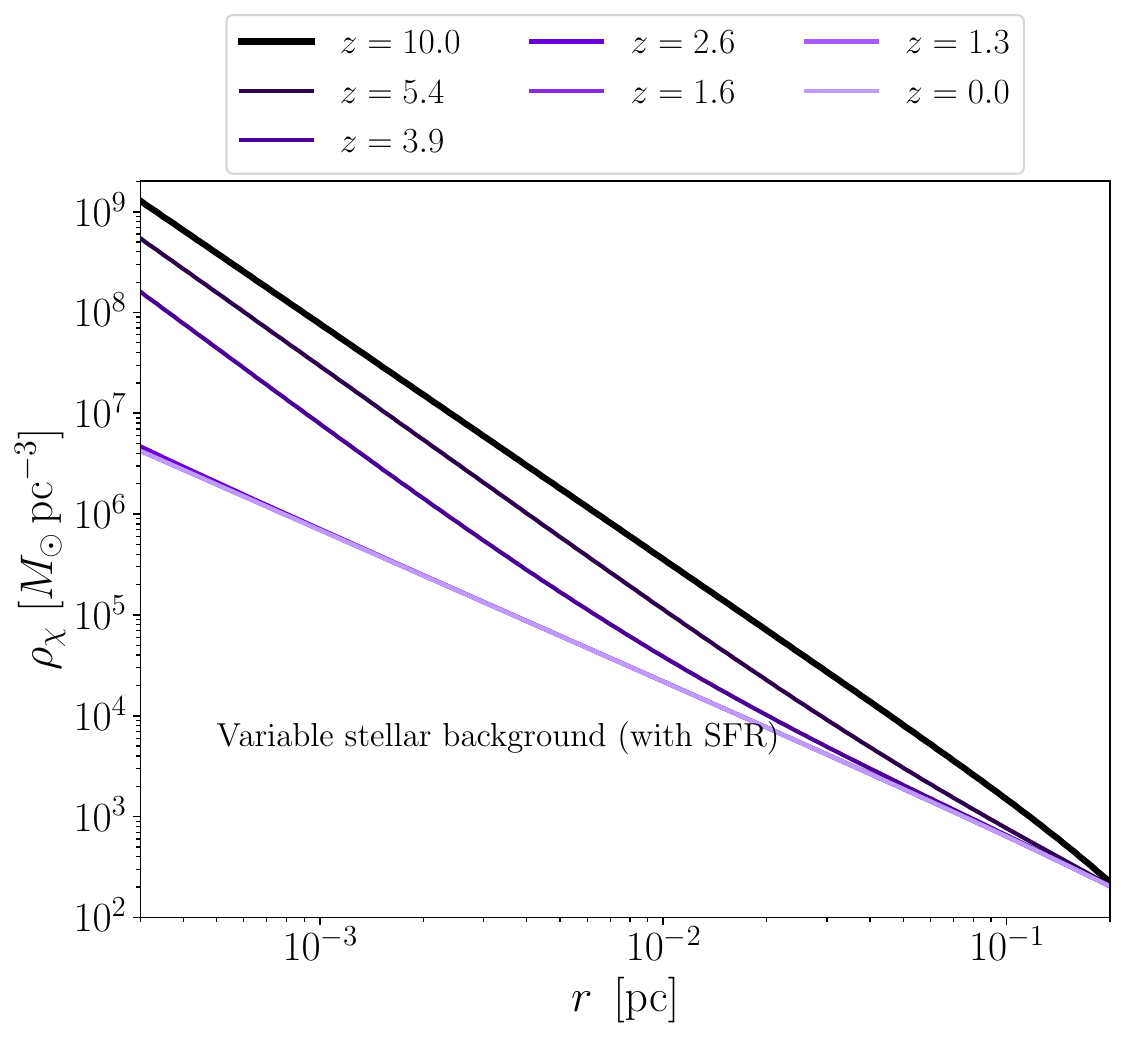}
  \caption{\textit{Upper left panel:} Phase space distribution of DM particles at different redshifts as a function of binding energy. The solid black line indicates the redshift of spike formation, and the colored blue lines indicate the accumulated effects of stellar heating at various redshifts. These lines assume a constant heating rate with parameters for the stellar profiles of $\rho_{\star}= 3.2 \times 10^5 M_{\odot} \mathrm{pc}^{-3}(r / 1 \mathrm{pc})^{-1}$ and $\gamma=1.4$ \cite{Genzel:2003cn}.  \textit{Upper right panel:} DM density profiles at various redshifts induced by stellar heating, consistent with the phase-space distributions shown in the left panel. \textit{Lower left panel:} Phase-space evolution of the DM at different redshifts, for a heating rate from stars scaling with the Star Forming Rate and normalization at $z=0$ fixed by parameters of $\rho_{\star}= 3.2 \times 10^5 M_{\odot} \mathrm{pc}^{-3}(r / 1 \mathrm{pc})^{-1}$ and $\gamma=1.4$ \cite{Genzel:2003cn}. \textit{Lower right panel:} DM density profile at various redshifts induced by stellar heating, and a heating rate scaling with redshift as the SFR.}
  \label{fig:stellar_heating}
\end{figure*}

\section{Cosmic Evolution of DM overdensities in a star cluster: continuous stellar bath}\label{sec:evolution_II}

\begin{figure*}[t!]
  \centering
\includegraphics[width=0.49\linewidth]{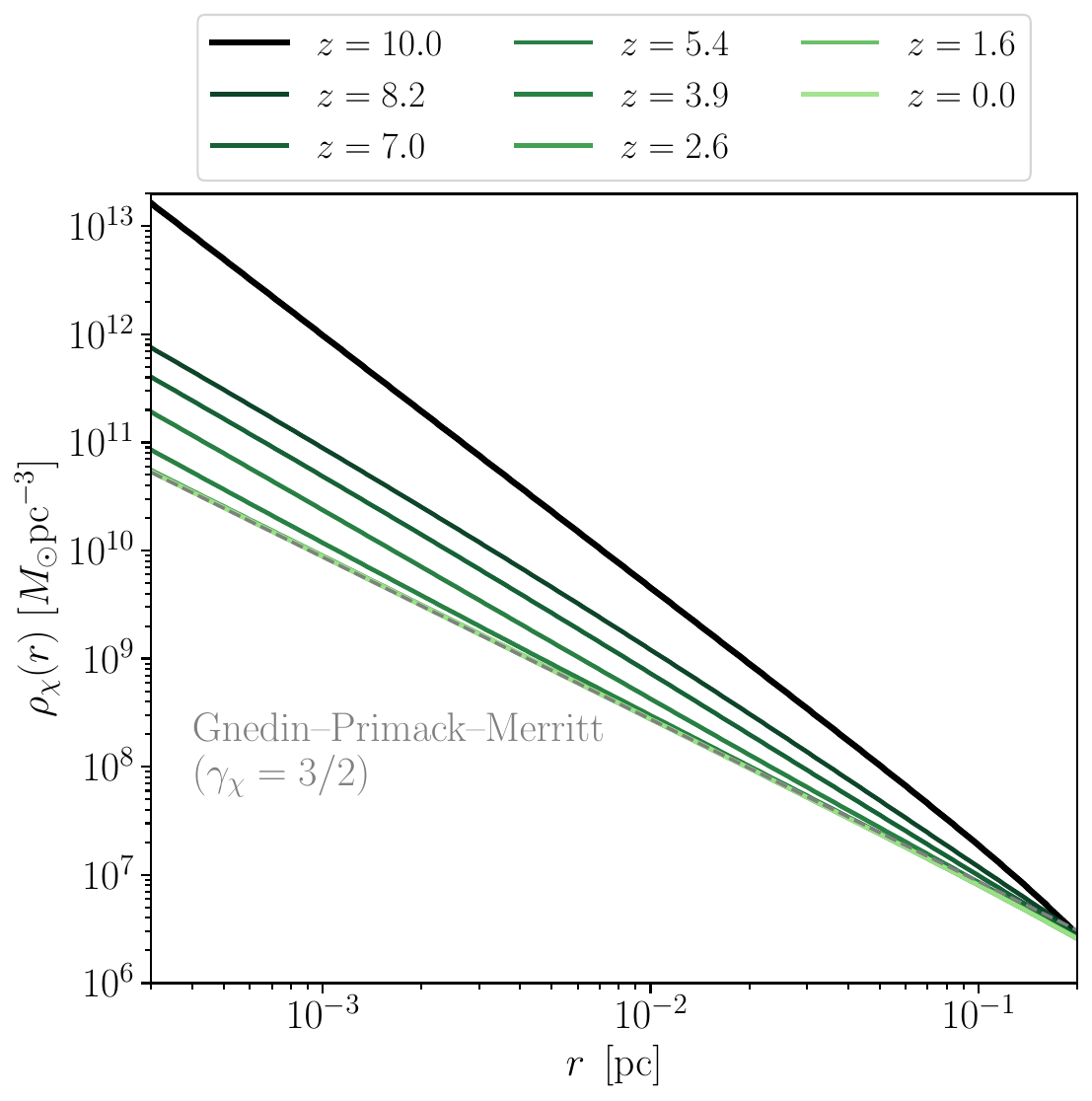}  \includegraphics[width=0.49\linewidth]{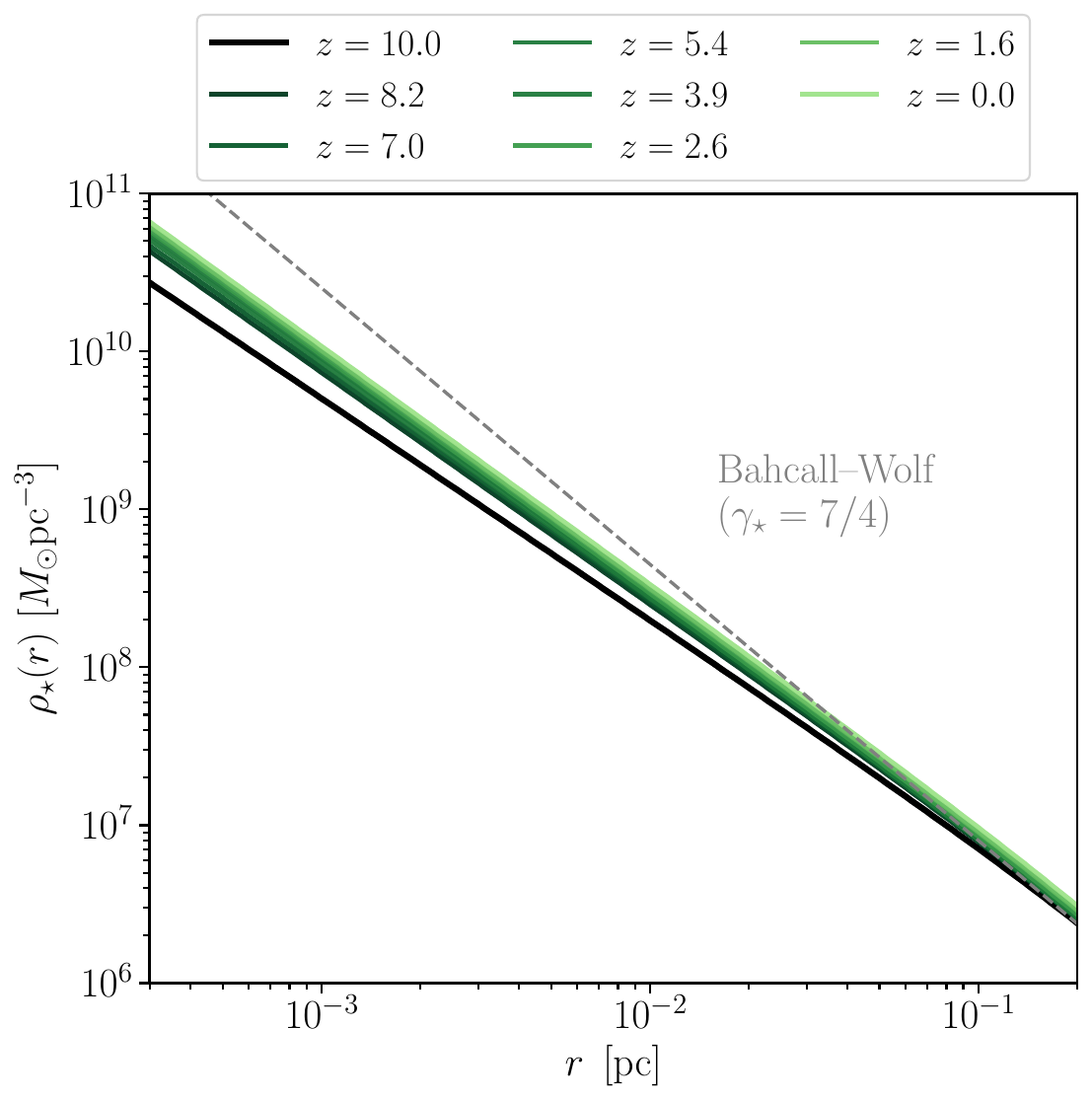}
  \caption{\textit{Left panel:}  Density profile of DM particles at various redshift stamps from the coevolution of the DM and stellar distributions. The steady state solution is shown as a dashed grey line. \textit{Right panel:} Density profile of stars at various redshift stamps from the coevolution of the DM and stellar distributions. The steady state solution for stars is shown as a dashed grey line.}
  \label{fig:coevolved_heating}
\end{figure*}
We begin by modelling a system analogous to that from Ref. \cite{Merritt:2003qk}. We solve the aforementioned system of Fokker-Planck equations while imposing physical boundary conditions on the system. We restrict the evolution to the region of gravitational influence of the black hole. In energy space, we bound the system as
\begin{equation}
\begin{aligned}
E &\in [E_h,\,E_{\max}], \ E_h \equiv \frac{G M_{\rm BH}}{r_h}, \\
E_{\max} &\equiv \frac{G M_{\rm BH}}{r_{\min}}, \,
r_{\min} \simeq 2R_S .
\end{aligned}
\end{equation}
with $r_{\rm min}$ denoting the black hole DM capture radius \cite{Gondolo:1999ef}. We further impose a reflecting outer boundary at \(E = E_h\), corresponding to the transition to the unperturbed outer halo extending beyond the spike radius. In the numerical implementation, this is enforced as a discrete equality \(f(E_0) = f(E_1)\), ensuring that the finite-difference flux term across the first energy cell vanishes. This is numerically equivalent to a Neumann condition \(\partial f / \partial E |_{E_h} = 0\). At the inner boundary, \(E = E_{\max}\), we impose an absorbing condition \(f(E_{\max}, t) = 0\), removing particles that have diffused onto tightly bound, capture-dominated orbits. These boundary conditions capture the two physical limits of the system. On the one hand, we assume that the outer halo remains a stationary reservoir on the relaxation timescale, which is well justified by the baryonic density falling steeply with radii. On the other hand, particles diffusing to the deepest potential region are accreted by the black hole. We note that a two-dimensional Fokker-Planck treatment in $(E,L)$ would capture loss-cone anisotropy; the preferential depletion of low-angular-momentum orbits that intersect the capture radius \cite{kohn_kulrsrud, 1977ApJ...211..244L, 1978ApJ...225..603S}. In the present isotropic framework, this effect is approximated by the absorbing inner boundary condition at $r_{\rm min} \simeq 2R_S$ \cite{Sadeghian:2013laa}. At radii sufficiently well above $r_{\rm min}$, the loss cone subtends a small solid angle in velocity space, and the fractional correction to the density from resolving the angular-momentum structure is correspondingly small. We also note that our reflecting (Neumann) outer boundary condition enforces a zero net flux across the influence radius. We expect this to be effectively equivalent, in the regime where the outer halo relaxation time greatly exceeds the inner one, to Monte Carlo Fokker-Planck schemes that replenish escaped particles from a Maxwell--Boltzmann distribution at the domain boundary \cite{1978ApJ...225..603S}.

We initialize the system with a canonical DM spike with index $\gamma_{\rm sp}=7/3$ \cite{Gondolo:1999ef, Quinlan:1994ed}, arising from an NFW profile with index $\gamma=1$, see Eq. \ref{eq:GSprofile}. The normalization of the spike is not particularly important for our task, as the system of Fokker-Planck equations will depend on the ratio of enclosed DM and stellar masses in the region of gravitational influence of the black hole, and as such we will assess our results in terms of that parameter. The size of the spike is of the order of the region of gravitational influence of the black hole for typical masses, see Eq. \ref{eq:spike_radius}.
We then map this density spike to energy space via Eddington inversion \cite{2008gady.book.....B}. For an isotropic system, the phase-space distribution function $f(\mathcal{E})$ can be recovered from the density profile $\rho(r)$ via
\begin{equation}
f(\mathcal{E})=
\frac{1}{\sqrt{8}\,\pi^2}
\left[
\int_0^{\mathcal{E}}
\frac{{\rm d}^2\rho}{{\rm d}\Psi^2}\,
\frac{{\rm d}\Psi}{\sqrt{\mathcal{E}-\Psi}}
+ \frac{1}{\sqrt{\mathcal{E}}}\,
\left(\frac{{\rm d}\rho}{{\rm d}\Psi}\right)_{\!\Psi=0}
\right],
\label{eq:eddington}
\end{equation}
where $\Psi(r)\equiv -\Phi(r)$ is the relative gravitational potential and $\mathcal{E}\equiv \Psi-\tfrac{1}{2}v^2$ the relative energy per unit mass. For a density cusp $\rho(r) \propto r^{-\gamma_{\rm sp}}$ embedded in a Keplerian potential $\Psi(r) = GM_{\rm BH}/r$, this can be solved analytically, yielding a power-law distribution function,
\begin{equation}
f(\mathcal{E}) = K\,\mathcal{E}^{\,\gamma_{\rm sp} - 3/2},
\end{equation}
with normalization constant \cite{Merritt:2003qk}
\begin{equation}
\begin{aligned}
K &= \frac{\rho_b\, r_b^{\gamma_{\rm sp}}}
{4\sqrt{2}\,\pi^{3/2}\,(G M_{\rm BH})^{\gamma_{\rm sp}}}
\,\frac{\Gamma(\gamma_{\rm sp}+1)}{\Gamma(\gamma_{\rm sp}-1/2)}
\\[4pt]
&= \frac{\rho_b\, r_b^{\gamma_{\rm sp}}}
{4\sqrt{2}\,\pi\,(G M_{\rm BH})^{\gamma_{\rm sp}}}\,
B\!\left(\gamma_{\rm sp}-\tfrac{1}{2},\,\tfrac{3}{2}\right),
\end{aligned}
\label{eq:K_beta}
\end{equation}
where $B(a,b) = \Gamma(a)\Gamma(b)/\Gamma(a+b)$ is the Euler Beta function. This normalization ensures that $f(\mathcal{E})$ reproduces the real-space density $\rho(r)$ upon integration over velocities. The resulting distribution defines the initial condition $f_\chi(\mathcal{E},t{=}0)$ used in the Fokker--Planck evolution, representing an unrelaxed dark-matter spike in dynamical equilibrium within the black hole potential.
At any time (or equivalently redshift) the configuration–space density follows from 
\begin{equation}
\begin{aligned}
\rho_\chi(r,t)
&= 4\sqrt{2}\,\pi
\int_{0}^{\Psi(r)} f(\mathcal{E},t)\,\sqrt{\Psi(r)-\mathcal{E}}\, d\mathcal{E}, \\
\Psi(r)
&= \frac{G M_{\rm BH}}{r}.
\end{aligned}
\label{eq:rho_from_f}
\end{equation}

\begin{figure*}[t!]
  \centering
\includegraphics[width=0.49\linewidth]{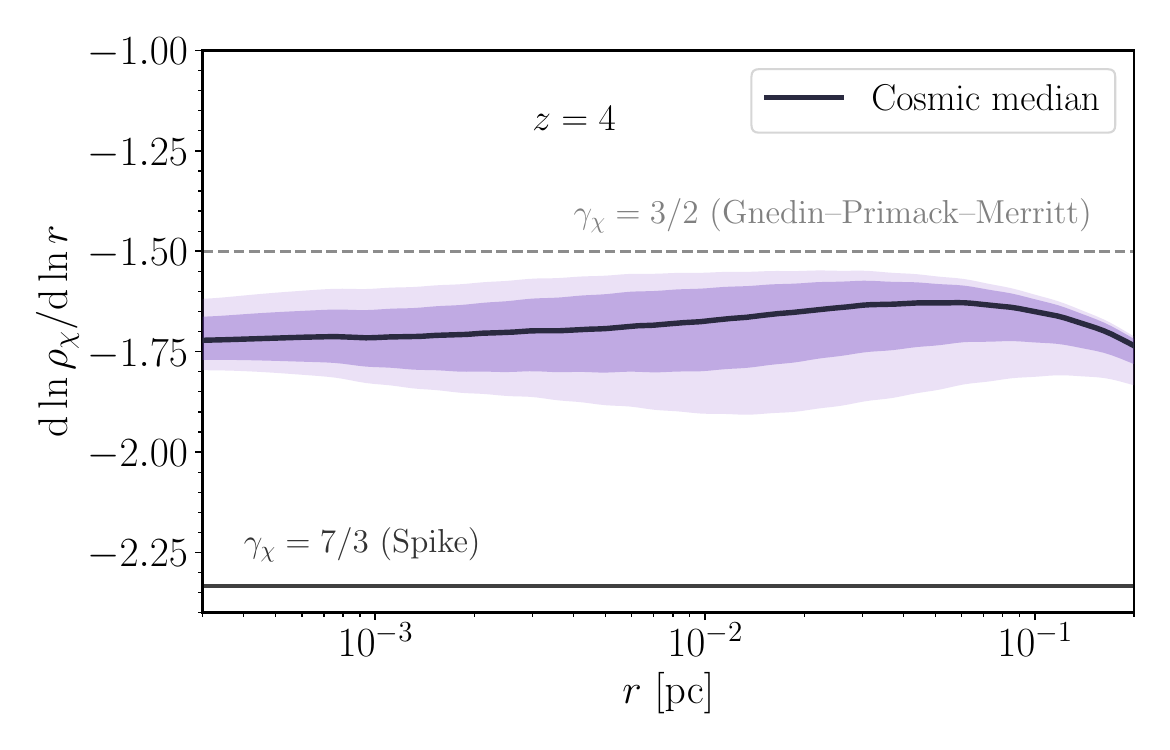} 
\includegraphics[width=0.49\linewidth]{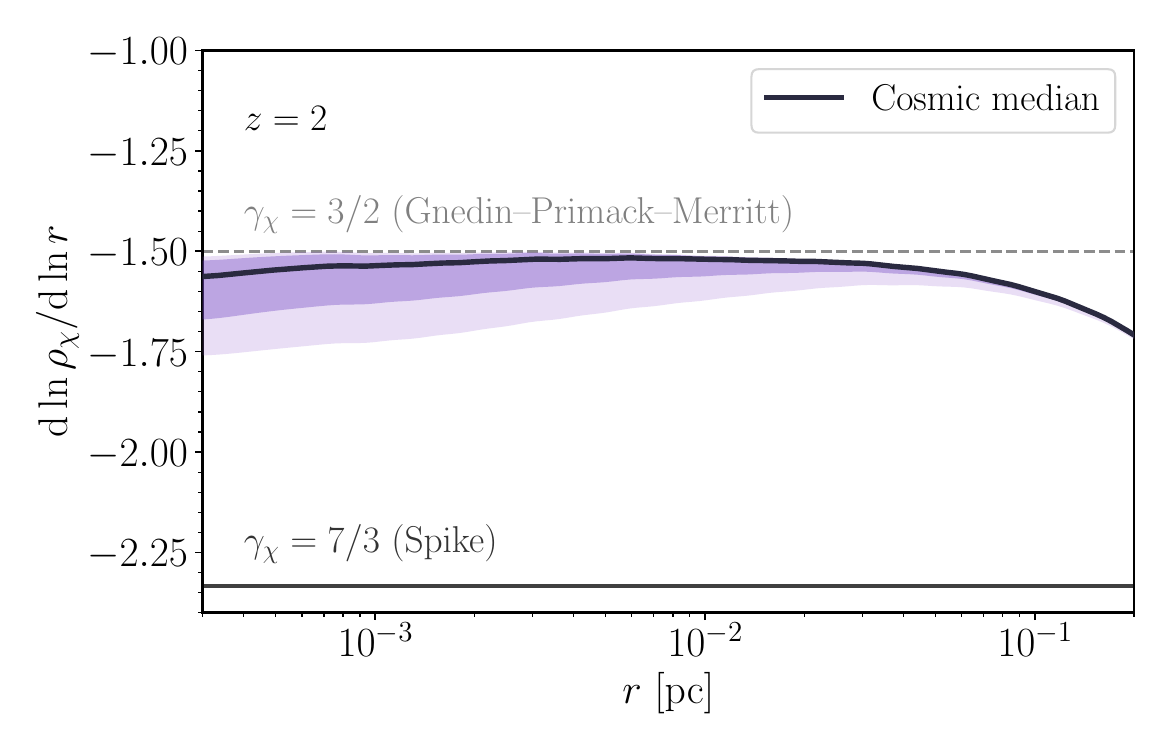} 
  \caption{Marginalized predictions for the DM logarithmic density slope
$s(r)\equiv \mathrm{d}\ln\rho_\chi/\mathrm{d}\ln r$, obtained from a two-component Fokker-Planck coevolution of DM and stars. The left (right) panel shows results at redshift $z=4$ ($z=2$), with the horizontal axis denoting radius and the vertical axis the local DM slope.
The solid black curve denotes the cosmic median profile, while the shaded regions indicate the $68\%$ and $95\%$ containment bands
after marginalizing over a three-dimensional parameter space:
the initial stellar cusp slope $\gamma_{\star,0}$,
the initial dark-matter spike slope $\gamma_{\chi,0}$,
and the enclosed mass ratio at the influence radius
$\eta \equiv M_\star(<r_h)/M_\chi(<r_h)$. We adopt physically motivated log-uniform priors with ranges
$\gamma_{\star,0}\in[1.2,1.8]$ \cite{1976ApJ...209..214B, Genzel_2010, Genzel:2003cn, Preto:2004kd},
$\gamma_{\chi,0}\in[2.1,2.3]$  \cite{Gondolo:1999ef, Ullio:2001fb},
and $\eta\in[0.3,3]$,
and sample this space using a Monte Carlo ensemble.
Horizontal reference lines indicate the canonical adiabatic spike ($\gamma_\chi=7/3$) and the Gnedin-Primack-Merritt attractor solution ($\gamma_\chi=3/2$).
At late times, the inner DM profile converges toward the latter, largely independent of initial conditions of the system.
}

  \label{fig:marginalized_slopes}
\end{figure*}
The rate of DM heating by stellar encounters is encoded implicitly in the diffusion coefficients. In particular, the coefficient \(D_{EE}\) is proportional to the mean squared change in binding energy per unit time, and scales as \(D_{EE} \propto G^2 M_\star^2 \rho_\star \ln\Lambda\). The overall normalization of the diffusion term can therefore be expressed in terms of a characteristic stellar-heating timescale \cite{Merritt:2003qk},
\begin{equation}
T_{\rm heat}
\simeq \frac{0.0814\,\sigma^3}{G^2 M_\star \rho_\star \ln\Lambda},
\label{eq:Theat}
\end{equation}
with \(\sigma\) denoting the stellar velocity dispersion. This quantity measures the timescale over which random gravitational encounters with stars redistribute energy in the dark--matter population. A smaller \(T_{\rm heat}\) corresponds to stronger heating and faster relaxation. 

It is convenient to work with a dimensionless time variable,
\begin{equation}
\tau \equiv \frac{t}{T_{\rm heat}},
\label{eq:tau_def}
\end{equation}
and to factor out the overall normalization of the diffusion  operator. In terms of \(\tau\), the Fokker-Planck equation for a fixed stellar bath can be written schematically as
\begin{equation}
4\pi^2 p(E)\,\frac{\partial f_\chi}{\partial \tau}
= \frac{\partial}{\partial E}
\left[
\tilde D_{EE}(E)\,\frac{\partial f_\chi}{\partial E}
\right],
\label{eq:FP_tau}
\end{equation}
where \(\tilde D_{EE}(E)\) is now a dimensionless diffusion coefficient of order unity that encodes the energy dependence, while the overall rate has been absorbed into \(\tau\). For a constant stellar background (fixed \(\rho_\star, M_\star, \sigma\)), the steady-state solution of Eq.~\ref{eq:FP_tau} has vanishing flux \(F_E=0\), leading to the classical Bahcall--Wolf cusp \(f_\star \propto E^{1/4}\) and \(\rho_\star \propto r^{-3/2}\) as the limiting configuration \cite{1976ApJ...209..214B}. This is illustrated in the upper panels of Fig. \ref{fig:stellar_heating}, where we show the evolution of the phase space distribution and density profile of an initial DM spike at $z=10$ due to stellar heating of a continuous stellar bath, with equal enclosed and DM masses at $r_h$, and stellar profile with index $\gamma_{\star}=1.5$. In the upper left panel, we note that the phase space distribution is suppressed more strongly with redshift at high binding energies, which translates in the density profiles being more discrepant in the innermost vicinity of the supermassive black hole. On the contrary, the phase space distributions tend to converge at low binding energies, which translates in comparable density profiles in outer radii.

The prescription used so far, with a continuous stellar bath, still does not account for the redshift-dependent evolution of the stellar component. We now generalize the diffusion rate to evolve with cosmic time. The local stellar density, and hence \(D_{EE}\), is expected to track the cosmic star-formation rate (SFR). We adopt the comoving SFR density from
\cite{Madau:2014bja},
\begin{equation}
\psi(z) = 0.015\,\frac{(1+z)^{2.7}}{1 + \left[(1+z)/2.9\right]^{5.6}}
\quad
M_\odot~\mathrm{yr^{-1}\,Mpc^{-3}},
\label{eq:sfr_madau14}
\end{equation}
and modulate the effective heating rate by the ratio \(\psi(z)/\psi(z{=}10)\).
Operationally, this amounts to replacing the constant timescale \(T_{\rm heat}\)
by a redshift--dependent effective heating time,
\begin{equation}
T_{\rm heat}(z) = T_{\rm heat}(z{=}10)\,
\frac{\psi(z{=}10)}{\psi(z)},
\end{equation}
or equivalently to evolving the system in terms of an ``effective'' dimensionless time
\begin{equation}
{\rm d}\tau_{\rm eff} = \frac{{\rm d}t}{T_{\rm heat}(z)}
= \frac{{\rm d}t}{T_{\rm heat}(z{=}10)}\,
\frac{\psi(z)}{\psi(z{=}10)}.
\end{equation}
In practice, we integrate the Fokker-Planck equation on a grid in redshift and multiply each time step by the factor \(\psi(z)/\psi(z{=}10)\), thereby accelerating diffusion at epochs of intense star formation (\(z\sim 2\)) and suppressing it at very early and late times. This is illustrated in the lower panels of Fig. \ref{fig:stellar_heating}. It can be noticed that this prescription causes DM spikes to relax more rapidly during the cosmic SFR peak, around $z\simeq 2-3$. As shown in Fig.~\ref{fig:stellar_heating}, including a SFR-dependent heating rate causes the spike to approach the Bahcall-Wolf configuration by \(z\lesssim 2\). With a constant heating rate, however, the spike is not fully depleted until \(z\lesssim 1\). In all cases we assume that the DM spike has
formed by \(z=10\).

\section{Coevolution of the DM and stellar distributions}
\label{sec:evolution_III}

In the previous section, we treated stars as a continuous bath with a static density profile index, allowing only the overall normalization to vary with the SFR when solving the Fokker-Planck equation. This prescription is adequate as long as the enclosed stellar mass exceeds the enclosed DM mass within the spike region. We now add another layer of realism by coevolving the DM and stellar density profiles in a self-consistent manner. This allows us to explore scenarios in which the enclosed DM mass is comparable to, or even larger than, the stellar mass. Following Ref. \cite{Merritt:2006mt}, the coupled Fokker-Planck equations read
\begin{align}
4\pi^{2} p(E)\,\frac{\partial f_\chi}{\partial t}
&= \frac{\partial}{\partial E}\!\left[D_{EE}^{\rm tot}(E,t)\,\frac{\partial f_\chi}{\partial E}\right],
\label{eq:FP_DM}\\[4pt]
4\pi^{2} p(E)\,\frac{\partial f_\star}{\partial t}
&= \frac{\partial}{\partial E}\!\left[D_{EE}^{\rm tot}(E,t)\,\frac{\partial f_\star}{\partial E}
\;+\;M_\star\,
D_E^{(\star)}(E,t)f_\star\right],
\label{eq:FP_star}
\end{align}
where the diffusion is sourced by all massive components,
\begin{equation}
\begin{aligned}
D_{EE}^{\rm tot}(E,t)
&= 64\pi^{4} G^{2}\ln\Lambda
\Bigg[
q(E)\!\int_{0}^{E} h(E',t)\,dE'
\\
&\quad
+ \int_{E}^{\infty} q(E')\,h(E',t)\,dE'
\Bigg], \\
h(E,t) &= m_\star f_\star + m_\chi f_\chi .
\end{aligned}
\label{eq:DEE_tot}
\end{equation}
and stars experience an additional drift term describing dynamical friction on the
stellar background,
\begin{equation}
D_E^\star(E,t)= \,64\pi^{4} G^{2}\ln\Lambda \int_{0}^{E}\! p(E')\,f_\star(E',t)\,dE'.
\label{eq:Astar}
\end{equation}
We can neglect the analogous drift term for dark matter, i.e.\ no \(D_E^\chi\), see Appendix \ref{sec:diffusion_coefficients} for a justification. As in the fixed-bath case, it is convenient to recast Eq.~\ref{eq:FP_DM} and Eq.
\ref{eq:FP_star} in terms of a dimensionless time variable. We define a reference heating timescale \(T_{\rm heat,0}\) using the initial stellar distribution, and introduce \(\tau = t/T_{\rm heat,0}\). The coupled equations then take the form
\begin{align}
4\pi^{2} p(E)\,\frac{\partial f_\chi}{\partial \tau}
&= \frac{\partial}{\partial E}\!\left[\tilde D_{EE}^{\rm tot}(E,\tau)\,\frac{\partial f_\chi}{\partial E}\right],
\\[4pt]
4\pi^{2} p(E)\,\frac{\partial f_\star}{\partial \tau}
&= \frac{\partial}{\partial E}\!\left[\tilde D_{EE}^{\rm tot}(E,\tau)\,\frac{\partial f_\star}{\partial E}
\;+\;M_\star\,
\tilde{D}_E^{(\star)}(E,\tau)f_\star\right],
\end{align}
where tildes denote diffusion and drift coefficients normalized by \(T_{\rm heat,0}\). In this formulation, the dimensionless time \(\tau\) directly measures the degree of relaxation of the system, while the evolving \(f_\chi\) and \(f_\star\) determine the instantaneous strength of diffusion via Eq.~\ref{eq:DEE_tot}. Compared to the fixed-bath case, this coevolution prescription introduces two genuinely new effects. First, the
diffusion strength \(D_{EE}^{\rm tot}\) now depends on both the stellar and DM distributions. Second, the stellar distribution function is no longer a static continuous bath, but also evolves towards a steady-state solution (the Bahcall--Wolf relaxation index \(\rho_\star\propto r^{-7/4}\)) at late times.

In Fig.~\ref{fig:coevolved_heating} we show the full coevolution of the dark-matter and stellar density profiles obtained by solving the coupled Fokker–Planck system in energy space, for a fiducial model with $M_\chi/M_\star=1$ inside the radius of influence. Starting from an initial configuration in which DM follows a steep adiabatic spike and stars a shallower cusp, stellar encounters drive both components towards their respective steady-state solutions. The left panel shows the dark-matter density profile at several redshift snapshots: the inner cusp relaxes from its initial spike-like slope to a profile close to $\rho_\chi\propto r^{-3/2}$. The right panel shows the corresponding stellar density, which converges towards the Bahcall–Wolf cusp $\rho_\star\propto r^{-7/4}$, as indicated by the grey dashed lines.
\begin{figure}[t!]
  \centering
\includegraphics[width=0.99\linewidth]{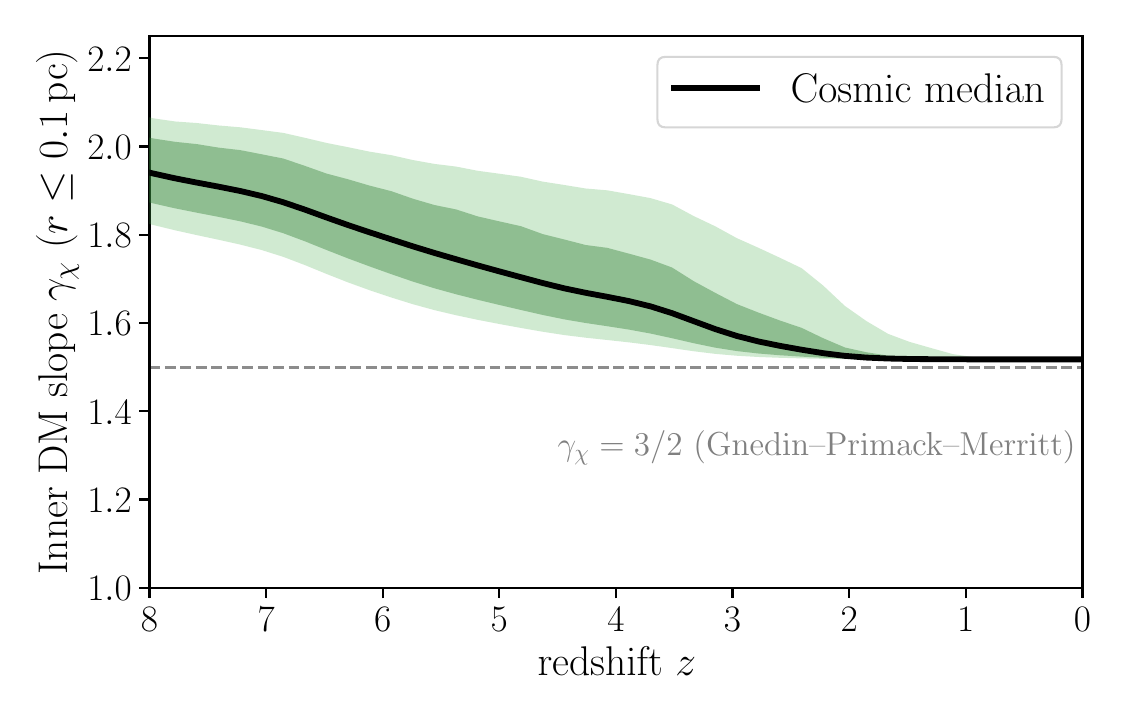} 
  \caption{Cosmic evolution of the averaged inner DM density slope
$\gamma_\chi$ within $r\leq 0.1\,\mathrm{pc}$. The black curve shows the median over a Monte Carlo ensemble of two-component
(stars + DM) Fokker-Planck evolutions, while shaded regions denote the
68\% and 95\% containment bands.
The ensemble marginalizes over log-uniform priors on the initial stellar cusp slope
$\gamma_{\star,0}\in[1.2,1.8]$, the initial dark-matter spike slope
$\gamma_{\chi,0}\in[2.1,2.3]$, and the enclosed mass ratio
$\eta=M_\star(<r_h)/M_\chi(<r_h)\in[0.3,3]$. At low redshifts, regardless of the initial conditions of the system, the inner slope converges to the steady state solution.
}
  \label{fig:cosmic_means}
\end{figure}

In Fig.~\ref{fig:marginalized_slopes} we show the marginalized prediction for the radial dependence of the dark-matter density slope,
\begin{equation}
\gamma_\chi(r) \equiv -\frac{{\rm d}\ln\rho_\chi}{{\rm d}\ln r},
\end{equation}
obtained from the same Monte Carlo ensemble of coupled Fokker--Planck evolutions described above. The left and right panels show the results at redshifts $z=4$ and $z=2$, respectively, after an initial spike has formed at $z_{\rm form}=10$. At $z=4$, the inner profile has already begun to relax but retains memory of the initial spike at intermediate radii, with a median slope that transitions from $\gamma_\chi\simeq 3/2$ at the smallest radii to values approaching the initial $\gamma_\chi\simeq 7/3$ at larger radii. The 68\% and 95\% containment bands reflect the spread in relaxation rates across the prior space. By $z=2$, the median profile has converged to $\gamma_\chi\simeq 3/2$ over a wider radial range, and the uncertainty bands narrow considerably, indicating that the system has largely lost memory of its initial conditions in the inner region. In Appendix~\ref{app:fixed_evolution}, we show the analogous results when varying individual parameters (enclosed mass ratio, stellar slope) at fixed values, which illustrate how each quantity separately affects the relaxation rate.

Figure~\ref{fig:cosmic_means} summarizes the global evolution of the inner cusp by showing the volume-averaged logarithmic slope $\langle\gamma_\chi\rangle$ within $r<0.1~{\rm pc}$ as a function of redshift, marginalized over the same prior space. The median inner slope (solid curve) decreases monotonically from the initial spike value towards the steady-state attractor $\gamma_\chi\simeq 3/2$, with the 68\% and 95\% containment bands bracketing the range of plausible evolutionary histories. Regardless of initial conditions, the median reaches $\langle\gamma_\chi\rangle\simeq 1.5$ by $z\lesssim 2$ for spikes formed at $z_{\rm form}=10$. The width of the bands reflects the sensitivity to the enclosed mass ratio and stellar cusp slope: systems with more massive or cuspier stellar components relax more quickly, while DM-dominated or shallow stellar configurations retain steeper cusps for longer. In Appendix~\ref{app:fixed_evolution}, we show the dependence on individual parameters (enclosed mass ratio, stellar slope, and the effect of a redshift-dependent SFR-weighted heating rate), illustrating how each quantity separately controls the relaxation rate and the transient behavior of the inner slope.

In Appendix \ref{app:convergence}, we discuss our numerical tolerance and discretization choices when solving the Fokker-Planck Eqs. \ref{eq:FP_master} and \ref{eq:FP_star}, and the ensuing impact on the simulated profiles. The impact on the simulated profile steepness evolutions is small, of less than $10 \%$, and the uncertainty on the normalizations of the profiles is somewhat larger, but in any case not larger than a factor of $\sim 2$ within $r=10^{-3}-10^{-1}$\,pc.

\begin{figure*}[t!]
  \centering
\includegraphics[width=0.69\linewidth]{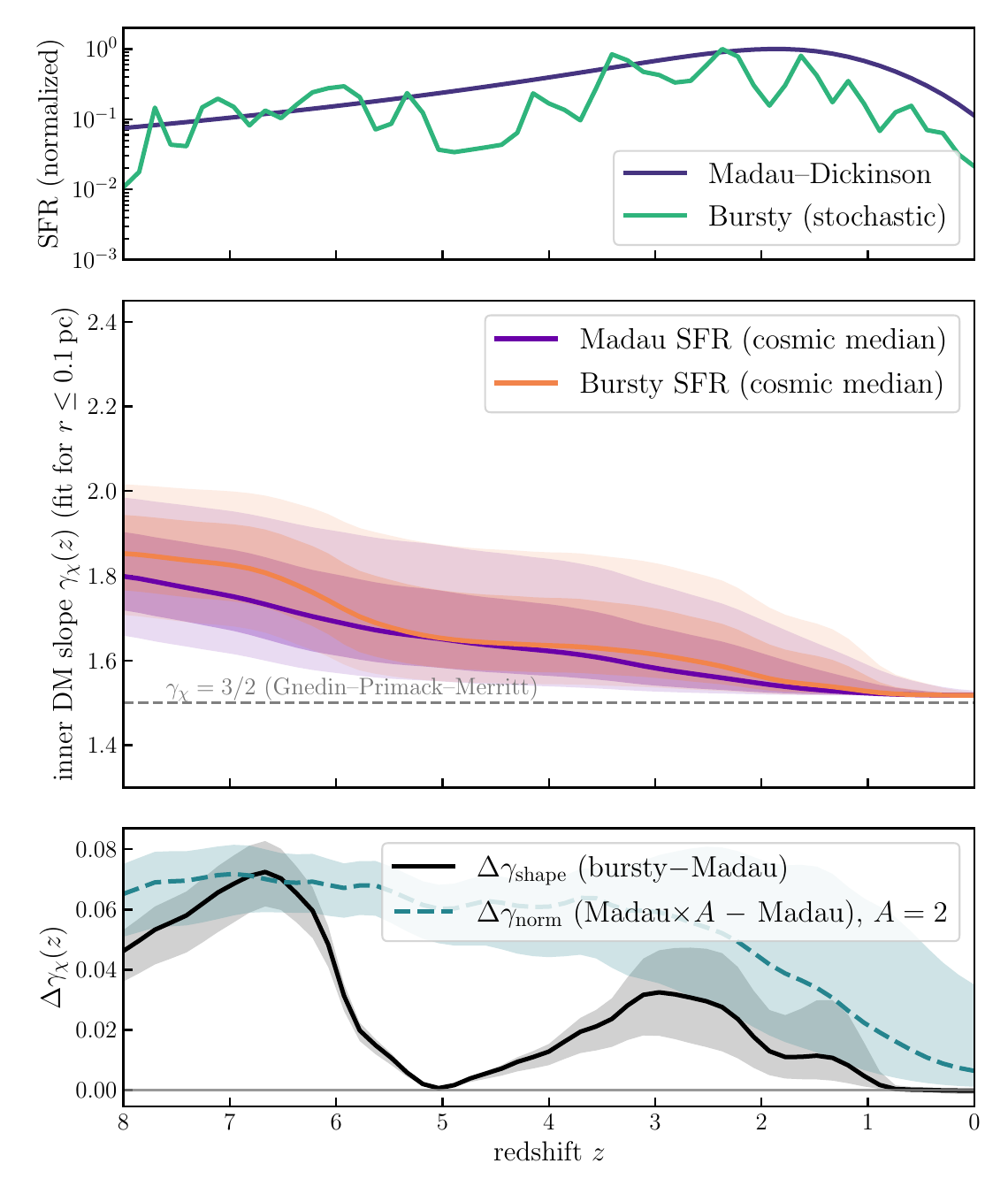} 
\caption{
Impact of the cosmic star forming rate on the evolution of the inner DM spike slope.
\textit{Top panel:} Comparison between a smooth star formation rate (SFR) \cite{Madau:2014bja} and a stochastic, bursty realization \cite{marszewski2026littlereddotsfire, Sparre_2016, 2020MNRAS.497..698T, 2023MNRAS.525.3254S}, normalized to equal integrated stellar mass over cosmic time.
\textit{Middle panel:} Redshift evolution of the averaged inner DM slope $\gamma_\chi(z)$ within  $r \le 0.1\,\mathrm{pc}$. Shaded bands indicate the 68\% and 95\% containment bands. The horizontal dashed line marks the attractor $\gamma_\chi = 3/2$ steady state solution.
\textit{Bottom panel:} Decomposition of the difference in spike slope into a pure shape effect (bursty minus smooth SFR at fixed normalization) and a normalization effect (rescaling of the smooth SFR amplitude). We find that temporal structure in star formation can modify the inner slope at the level of 10-20$\%$, comparable to overall normalization shifts. A larger but still mild effect is observed when enhancing the normalization of the SFR smooth shape by a factor of 2.
}
  \label{fig:sfr_bursty}
\end{figure*}
\section{Impact of Star Forming Rate}\label{sec:sfr_impact}

In some galaxies, the star-formation history (SFH) may deviate substantially from the smooth, cosmic-average star-formation rate density parametrizations (e.g. \cite{Madau:2014bja}). In particular, both observations of resolved stellar populations in nearby dwarfs and feedback-regulated galaxy-formation models indicate that star formation can proceed in intermittent episodes, with phases of enhanced activity separated by relatively quiescent periods (``bursty'' star formation) \cite{Weisz:2014yfa, pan2023modellingstochasticstarformation, marszewski2026littlereddotsfire}. 

To assess the impact of such temporal variability on the dynamical evolution of the dark-matter distribution, we implement a phenomenological bursty SFH by modulating a baseline, smooth SFH with a dimensionless burst factor,
\begin{equation}
\dot{M}_\star^{\rm bursty}(z)\;=\;\mathcal{N}\,\dot{M}_\star^{\rm smooth}(z)\,\Big[1+\sum_{i=1}^{N_{\rm b}} A_i\,\exp\!\Big(-\frac{(z-z_i)^2}{2\sigma_z^2}\Big)\Big]\,,
\end{equation}
where $A_i$ controls the burst amplitude, $z_i$ the burst redshift, and $\sigma_z$ its duration in redshift space. The prefactor $\mathcal{N}$ is fixed by requiring that the bursty and smooth models form the same total stellar mass over the redshift, so that the comparison isolates the effect of temporal redistribution of star formation (burstiness) at fixed integrated stellar mass formed. The impact of a bursty SFR versus a smooth parametrization in the DM spike slope is shown in Fig \ref{fig:sfr_bursty}. When varying the shape of the SFR but retaining a consistent normalization after integrating over redshifts, the effects on the spike slope evolution are mild, of less than $15\%$, and only manifest at certain redshifts. When varying the normalization of the smooth parametrization by a factor of 2, we find a slightly larger effect, of $20\%$, and more homogeneously distributed over redshift.

\section{Impact of spike formation redshift}\label{sec:formation_z}

Throughout the preceding sections we have assumed that the DM spike forms at $z_{\rm form}=10$. In practice, the formation redshift depends on when the central black hole first reaches the mass required for adiabatic growth to produce a significant overdensity. For black holes that assemble their mass more gradually or in later-forming halos, the spike may be established at considerably lower redshifts, leaving less cosmic time for stellar heating to soften the inner profile before the present epoch. To quantify this effect, we repeat the coupled two-component Fokker--Planck evolution of Sec.~\ref{sec:evolution_III} for four representative formation redshifts, $z_{\rm form}\in\{10,\,8,\,6,\,4\}$, keeping all other model parameters fixed and using the same Monte Carlo parameter draws for each $z_{\rm form}$.

\begin{figure}[t!]
  \centering
\includegraphics[width=0.99\linewidth]{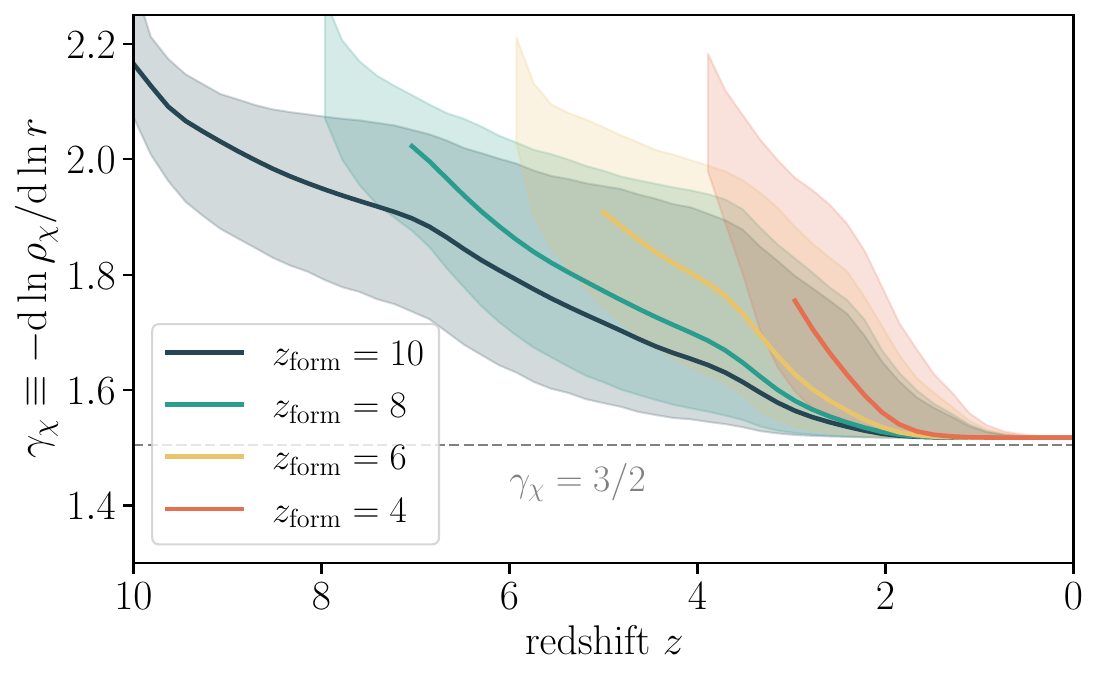}
\caption{Impact of the assumed spike formation redshift on the evolution of the inner DM density slope $\gamma_\chi \equiv -\mathrm{d}\ln\rho_\chi/\mathrm{d}\ln r$, averaged within $r \le 0.1\,\mathrm{pc}$. Each colored curve shows the median redshift evolution for a different formation redshift $z_{\rm form}\in\{10,\,8,\,6,\,4\}$, obtained from a Monte Carlo ensemble of coupled two-component Fokker--Planck evolutions marginalizing over the initial stellar cusp slope $\gamma_{\star,0}\in[1.2,1.8]$, the initial spike slope $\gamma_{\chi,0}\in[2.1,2.3]$, and the enclosed mass ratio $\eta\in[0.3,3]$ (darker shaded bands, 68\% containment). The lighter background bands show the 68\% interval for a broader, conservative prior choice. The horizontal dashed line marks the steady-state attractor $\gamma_\chi=3/2$. Regardless of when the spike forms, the system converges toward this attractor, but later formation shifts the relaxation to lower redshifts.}
\label{fig:formation_redshift}
\end{figure}

The results are shown in Fig.~\ref{fig:formation_redshift}. At each $z_{\rm form}$, the inner slope starts near the canonical spike value $\gamma_\chi\simeq 7/3$ and subsequently relaxes towards the steady-state attractor $\gamma_\chi\simeq 3/2$, as expected from stellar gravitational heating. The key difference is the redshift at which this relaxation is completed: spikes formed at $z_{\rm form}=10$ have converged to the steady state by $z\simeq 2$, whereas spikes formed at $z_{\rm form}=4$ retain significantly steeper inner slopes ($\gamma_\chi \gtrsim 2$) at $z\simeq 2$, and may not have fully relaxed even by $z=0$. For all formation redshifts, the qualitative evolution is the same, and the steady-state solution remains a robust attractor. However, the quantitative predictions for the inner slope at any given observational redshift depend on when the spike was established. This implies that younger galactic nuclei, or black holes that grew to their final mass at lower redshifts, may harbor steeper DM overdensities than older systems with comparable final masses.

\section{Implications for particle DM searches}\label{sec:implications}

The redshift evolution of the central column density,
\begin{equation}
\Sigma(z) = 2\int_{r_{\min}}^{r_{\max}} \rho_\chi(r,z)\,dr,
\end{equation}
and of the central annihilation proxy,
\begin{equation}
J(z) = 2\int_{r_{\min}}^{r_{\max}} \rho_\chi^2(r,z)\,dr,
\end{equation}
has implications for indirect and direct probes of particle DM. Because different particle-physics scenarios depend on different powers of the dark-matter density, the progressive depletion of the inner spike affects annihilation, decay, boosting and scattering searches in qualitatively distinct ways.

\begin{figure}[t!]
  \centering
\includegraphics[width=0.99\linewidth]{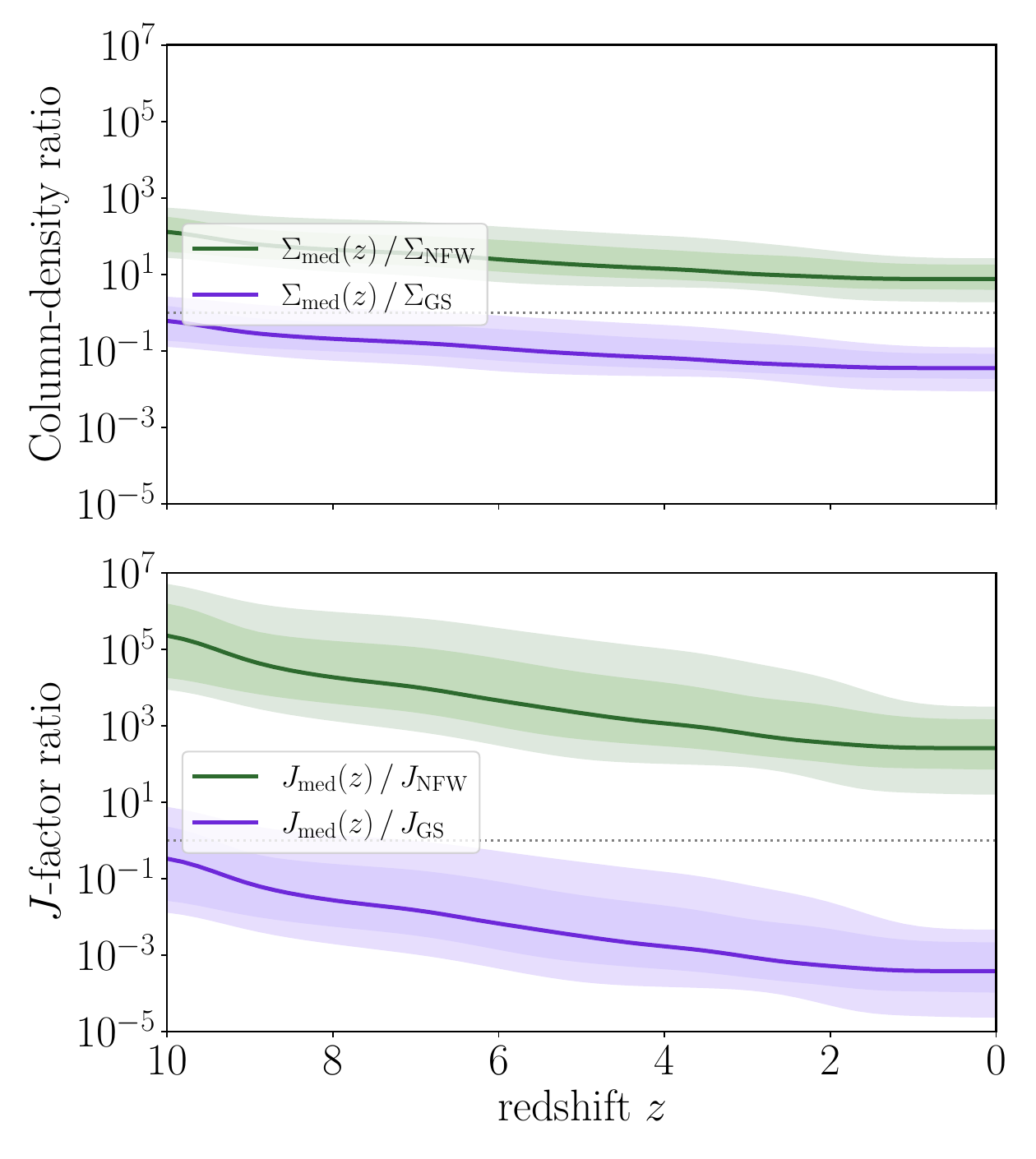} 
\caption{Redshift evolution of the central dark-matter column density and $J$-factor ratios, integrated within $r_{\max} = 0.1\,\mathrm{pc}$.
In each panel, the green curve shows the ratio of the Monte Carlo cosmic median to the NFW expectation,
$\Sigma_{\rm med}(z)/\Sigma_{\rm NFW}$ and $J_{\rm med}(z)/J_{\rm NFW}$ (above unity),
while the purple curve shows the ratio of the cosmic median to the canonical Gondolo-Silk prediction,
$\Sigma_{\rm med}(z)/\Sigma_{\rm GS}$ and $J_{\rm med}(z)/J_{\rm GS}$ (below unity). The evolved spike remains denser than an uncontracted NFW cusp, but falling short of the canonical adiabatic spike.
Shaded regions indicate the 68\% (dark band) and 95\% (light band) containment intervals obtained from marginalization over the formation and. The redshift axis runs from high redshift (left) to the present epoch (right).
}
  \label{fig:implications}
\end{figure}

For self-annihilating DM, the photon or neutrino flux scales as
\begin{equation}
\Phi_{\rm ann} \propto \langle \sigma v \rangle \int \rho_\chi^2 \, d\ell \, d\Omega,
\end{equation}
i.e.\ with the square of the density. As a consequence, annihilation signals are quite sensitive to the inner slope of the density profile. Canonical adiabatic spikes with $\gamma_{\rm sp}=7/3$ predict very large central $J$-factors, potentially enhancing annihilation signals by orders of magnitude relative to an NFW cusp. Our results show, however, that stellar heating and long-term dynamical evolution drive the system toward a shallower, quasi–steady-state configuration, reducing the central $J$-factor over cosmic time. This depletion can significantly weaken annihilation signals relative to naive canonical-spike expectations, concretely about 2 to 4 orders of magnitude, see Fig. \ref{fig:implications}.

For decaying DM scenarios, the signal scales linearly with the density,
\begin{equation}
\Phi_{\rm dec} \propto \Gamma \int \rho_\chi \, d\ell \, d\Omega,
\end{equation}
and is therefore controlled by the column density rather than by $\rho^2$. The dependence on the inner slope is milder in this case, of about 1 to 2 orders of magnitude, see Fig. \ref{fig:implications}.

In scenarios where DM scatters off Standard Model particles, the signal typically scales as
\begin{equation}
\Phi_{\rm scat} \propto \sigma_{\chi X} \int \rho_\chi \, n_X \, d\ell,
\end{equation}
where $n_X$ is the number density of the target species (e.g.\ gas, radiation fields, or cosmic rays). In such cases, the relevant astrophysical factor is again linear in $\rho_\chi$, implying a dependence similar to that of decaying DM. 

However, in environments where both DM and the target population are centrally concentrated, the effective signal can acquire a steeper radial weighting, partially restoring sensitivity to the inner slope. The redshift evolution of the spike therefore impacts scattering-induced gamma-ray or neutrino fluxes in an intermediate way. Less dramatically than annihilation signals, but potentially more strongly than pure decay scenarios depending on the spatial distribution of the target species.

An important consequence of the evolution we find is that spike-induced enhancements in the column density and $J$-factor are largest at high redshift and progressively reduced toward the present epoch, as shown in Fig.~\ref{fig:implications}. For the column density, the ratio $\Sigma_{\rm med}/\Sigma_{\rm NFW}$ decreases by roughly one order of magnitude between $z=10$ and $z=0$, while the ratio $\Sigma_{\rm med}/\Sigma_{\rm GS}$ drops from near unity to $\sim 10^{-2}$. The $J$-factor, being quadratic in the density, is far more sensitive. $J_{\rm med}/J_{\rm NFW}$ decreases by about 3 orders of magnitude over the same interval, and $J_{\rm med}/J_{\rm GS}$ falls to $\sim 10^{-4}$ by the present epoch. Redshift-dependent probes of the DM microphysics would then provide a path to distinguish from other astrophysical backgrounds, and here we provide an educated estimate of the redshift evolution of the size of the DM induced signals.

\section{Conclusions}\label{sec:conclusions}
Here we have discussed the formation and cosmic evolution of DM overdensities around massive black holes embedded in more realistic stellar environments. Our goal was to bridge the gap between idealized adiabatic spike models and the more complex conditions expected in galactic nuclei and dense star clusters, and to provide a physically grounded estimate of the DM inner slopes that are likely to be realized at various cosmological times. These predictions are relevant for a number of possible phenomenological observables of DM, such as its impact on multi-messenger emissions, see \textit{e.g} \cite{Herrera:2023nww, Cline:2022qld, Ferrer:2022kei,Herrera:2025gpm,Gondolo:1999ef, Bertone:2024rxe, Wang:2021jic}.

On the formation side, we generalized the standard Gondolo and Silk prescription to include a non-zero initial black hole seed mass, a surrounding stellar cusp with arbitrary normalization and slope, and non-circular DM orbits treated via full radial-action conservation. We showed that both the initial stellar distribution and the seed mass have a sizable impact on the resulting overdensity. Light seeds and low stellar enclosed masses allow spikes that approach the canonical $\gamma_{\rm sp}=7/3$ expectation, whereas heavier seeds and/or dense stellar cusps lead to substantially softer inner profiles already at formation. By marginalizing over the uncertain initial conditions, we found that the circular-orbit adiabatic contraction already produces softer inner slopes than the canonical prediction, and that accounting for non-circular orbits via radial-action conservation yields slightly steeper profiles in the innermost radii.

We then followed the subsequent evolution of these overdensities in energy space by solving orbit-averaged Fokker-Planck equations. First, we considered a fixed stellar bath, both with constant normalization and with a heating rate that tracks the cosmic star-formation history. In both cases, the DM approaches a steady-state solution, with $f_\chi(E)\to \mathrm{const}$ and $\rho_\chi(r)\propto r^{-3/2}$ inside the radius of influence, in agreement with \cite{Merritt:2003qk, Vasiliev:2008uz}. The rate at which this limit is reached is controlled primarily by the enclosed stellar-to-DM mass ratio and the steepness of the stellar cusp. For typical stellar environments, initial spikes formed at $z=10$ relax to $\gamma_\chi\simeq 3/2$ by $z\sim 2$, especially when the diffusion rate is enhanced around the SFR peak. In more dark-matter–dominated or shallower stellar systems, the relaxation is slower and the spike-like slopes can persist to lower redshifts, though they are ultimately driven towards the same asymptotic value.

We then relaxed the fixed-bath assumption and solved the coupled Fokker-Planck system for stars and DM simultaneously. In this coevolution setup, both components respond to a common diffusion operator, with stars additionally experiencing a drift term. We found that once the enclosed DM mass becomes comparable to the stellar mass, the two populations no longer evolve independently: DM relaxes towards $\rho_\chi\propto r^{-3/2}$ while stars converge to the Bahcall–Wolf cusp $\rho_\star\propto r^{-7/4}$, and the characteristic transition radii are very similar for both components. This confirms that the steady-state solution derived analytically in the classic Bahcall–Wolf and Gnedin–Primack–Merritt analyses is a robust attractor of the system, even when starting from steep adiabatic spikes and allowing for realistic stellar profiles and finite seed masses.

We then conclude that in realistic star clusters with non-zero black hole seeds and stellar cusps, the adiabatic growth phase still generically produces significant DM overdensities. However, the resulting profiles at formation are sizably less dense than the Gondolo–Silk spikes once the stellar mass and seed mass are taken into account. Gravitational encounters with stars efficiently erase the memory of the initial spike shape in the inner region, and do so fairly quickly in cosmic time. For a wide range of initial conditions and enclosed mass ratios, DM cusps formed at high redshift relax towards an inner slope $\gamma_\chi \simeq 1.5$ by $z\lesssim 2$, particularly if the stellar density follows the cosmic star-formation history.

These findings have several implications for indirect and direct DM searches and for dynamical probes of galactic nuclei. First, they indicate that extrapolations based on the steepest adiabatic-spike models likely overestimate the DM density near massive black holes by orders of magnitude at low redshift, especially in systems with significant stellar content. Second, they suggest that a $\gamma_\chi\simeq 1.5$ cusp is a more realistic benchmark for computing annihilation signals, scattering rates with high-energy messengers, boosted DM fluxes at direct detection experiments, or modifications to gravitational waveforms from compact-object inspirals in DM–rich environments. At the same time, the overdensities we find remain systematically cuspier than a canonical NFW-like profile, leaving ample room for enhanced DM phenomenological signatures.

Our work has caveats and can be improved in several directions. For instance, we have not included the impact of black hole mergers, the AGN gas component, repeated episodes of strong AGN feedback, mass segregation among multiple stellar populations, or departures from spherical symmetry, 
{i.e the galactic bar \cite{1991ApJ...379..631B}}. Nor have we modeled a full distribution of black hole seed masses and formation redshifts, which would broaden the predicted distribution of inner slopes across the galaxy population. Further, we have adopted a single representative stellar mass throughout this work. In practice, mass segregation in nuclear star clusters produces a mass-dependent cusp structure, with heavier stars dominating the heating rate at small radii, which could modify the relaxation timescales quantitatively. One may further include relativistic corrections in the formation part of our study, which have been shown to induce departures in the distribution very close to the Schwarzschild radii \cite{Sadeghian:2013laa,Ferrer:2017xwm,Caiozzo:2025mye}. Additionally, sizable DM self-interactions or DM-baryon interactions may impact the Fokker-Planck treatment employed. We leave these extensions to future work.

Nonetheless, the picture that emerges from our calculations is simple and robust. In realistic stellar nuclei, DM spikes are not expected to be fully erased nor as extreme as in the most optimistic adiabatic models. Instead, they tend to settle into a characteristic $\gamma_\chi\simeq 1.5$ cusp within a few Gyr, ($z\lesssim 2$ for spikes formed at $z\simeq10$), providing a well-motivated target for theoretical predictions and observational searches that aim to use galactic centers and star clusters as laboratories for the particle nature of DM. 

More broadly, the central lesson of this work is that DM spikes are not static structures, instead, their properties evolve significantly over cosmic time. A future robust detection of the microphysical nature of DM will likely rely on astrophysical observations at multiple redshifts. Here we have provided a framework and an educated estimate of the redshift evolution of the DM distribution in the vicinity of black holes.
\section{Acknowledgments}
We are grateful to Anna-Christina Eilers, Kohta Murase, Zeineb Mezghanni and Nathaniel Starkman for useful discussions. The work of GH is supported by the Neutrino Theory Network Fellowship with contract number 726844. This manuscript has been authored by FermiForward Discovery Group, LLC under Contract No. 89243024CSC000002 with the U.S. Department of Energy, Office of Science, Office of High Energy Physics. AH acknowledges that this material is based upon work supported by the U.S. Department of Energy, Office of Science, Office of Workforce Development for Teachers and Scientists, Office of Science Graduate Student Research (SCGSR) program. The SCGSR program is administered by the Oak Ridge Institute for Science and Education (ORISE) for the DOE. ORISE is managed by ORAU under contract number DESC0014664. All opinions expressed in this paper are the author’s and do not necessarily reflect the policies and views of DOE, ORAU, or ORISE.
L.N. is supported by the Sloan Fellowship and the NSF CAREER award AST-2337864.
\bibliography{references}

\appendix
\clearpage
\onecolumngrid

\section{DM spikes à la Gondolo and Silk}\label{app:GS_derivation}
Gondolo and Silk adopted the formalism applied for baryons by previous studies, and derived the distribution of DM particles around the central black hole for different assumptions on the inner slope of the DM halo and its phase-space distribution \cite{Gondolo:1999ef}.

Gondolo and Silk considered models of initial profiles with an inner core and an inner cusp. In both cases the system of equations to solve is given by the conservation of the integrals of motion:
\begin{equation}\label{eq:rho_integral_GS}
\rho_{f}(r)=\int_{E_f^{m}}^0 d E_{f} \int_{L_m^{c}}^{L_f^{m}} d L_{f} \frac{4 \pi L_{f}}{r^2 v_r} f_{f}\left(E_{f}, L_{f}\right)
\end{equation}

\begin{equation}
v_r  =\left[2\left(E_{f}+\frac{G M}{r}-\frac{L_{f}}{2 r^2}\right)\right]^{1 / 2} ,
\end{equation}
and
\begin{equation}
E_f^{m}  =-\frac{G M}{r}\left(1-\frac{4 R_{\mathrm{S}}}{r}\right),
\end{equation}
\begin{equation}
L_f^{c}  =2 c R_{\mathrm{S}},
\end{equation}
\begin{equation}
L_f^{m}  =\left[2 r^2\left(E_{f}+\frac{G M}{r}\right)\right]^{1 / 2} .
\end{equation}

which is derived from the adiabatic conditions
\begin{equation}
f\left(E_{f}, L_{f}\right)=f(E, L), L_{f}=L, I\left(E_{f}, L_{f}\right)=I(E, L)
\end{equation}
corresponding to conservation of the phase-space distribution, angular momentum and radial action of the DM particles.

For models with an inner cusp \cite{Navarro:1995iw, Navarro:1996gj}, $\rho(r)=$ $\rho_0\left(r / r_0\right)^{-\gamma}$, with $r_0$ the scale radius of the galaxy, and the phase-space distribution reads
\begin{equation}
f(E, L)=\frac{\rho_0}{\left(2 \pi \phi_0\right)^{3 / 2}} \frac{\Gamma(\beta)}{\Gamma\left(\beta-\frac{3}{2}\right)} \frac{\phi_0^\beta}{E^\beta},
\end{equation}
with $\beta=(6-\gamma) /[2(2-\gamma)]$ and $\phi_0=4 \pi G r_0^2 \rho_0 /[(3-\gamma)(2-$ $\gamma)$ ]. Assuming a potential proportional to $r^{2-\gamma}$ at small radii, the action integral cannot be performed exactly. Gondolo and Silk found an approximation good to better than $8 \%$ over all of phase space for $0<\gamma<2$ :
\begin{equation}
I(E, L)=\frac{2 \pi}{b}\left[-\frac{L}{\lambda}+\sqrt{2 r_0^2 \phi_0}\left(\frac{E}{\phi_0}\right)^{\frac{4-\gamma}{2(2-\gamma)}}\right],
\end{equation}
where $\lambda=[2 /(4-\gamma)]^{1 /(2-\gamma)}[(2-\gamma) /(4-\gamma)]^{1 / 2}$ and $b=$ $\pi(2-\gamma) / \mathrm{B}\left(\frac{1}{2-\gamma}, \frac{3}{2}\right)$. Expressing $E$ as a function of $E_{f}$ and integrating \ref{eq:rho_integral_GS}, they obtain
\begin{equation}\label{eq:profile_GS}
\rho_{\rm sp}(r) \equiv \rho_{f}(r)=\rho_R g_\gamma(r)\left(\frac{R_{\mathrm{sp}}}{r}\right)^{\gamma_{\mathrm{sp}}}
\end{equation}
with $\rho_R=\rho_0\left(R_{\mathrm{sp}} / r_0\right)^{-\gamma}, \gamma_{\mathrm{sp}}=(9-2 \gamma) /(4-\gamma)$ is the cuspiness of the spike, and the size of the spike is
\begin{equation}\label{eq:spike_radius_GS}
R_{\mathrm{sp}}=\alpha_\gamma r_0\left(M / \rho_0 r_0^3\right)^{1 /(3-\gamma)}.
\end{equation}
where $\alpha_{\gamma}$ is a normalization factor and $g_{\gamma}(r)$ accounts for the particles captured by the black hole, and  can be approximated for $0<\gamma <2 $ by  $g_{\gamma}(r) \simeq (1-\frac{4R_{\rm S}}{r})^3$, with $R_{\rm S}$ the Schwarzschild radius. In particular, $\alpha_\gamma \simeq 0.293 \gamma^{4 / 9}$ for $\gamma \ll 1$, and is $\alpha_\gamma=0.00733$, $0.120,0.140,0.142,0.135,0.122,0.103,0.0818,0.0177$ at $\gamma=0.05,0.2,0.4, \ldots, 1.4,2$. The density falls abruptly to zero at $r \lesssim 10 R_{\mathrm{S}}$, and vanishes for $r<4 R_{\mathrm{S}}$, which is however a conservative assumption that neglects relativistic and rotating effects in black holes into account \cite{Sadeghian:2013laa, Ferrer:2017xwm}. We notice that for $0 < \gamma < 2$, the density slope in the spike, $\gamma_{\mathrm{sp}}$, varies only between 2.25 and 2.5. The slope of the spike profile from equation \ref{eq:profile_GS} is consistent with the result from Ref. \cite{Quinlan:1994ed}, where models with $0 <\gamma < 2$ obey
\begin{equation}
n=\frac{6-\gamma}{2(2-\gamma)}.
\end{equation}
with the exception of $\gamma=0$. In this case, $n=1$ and the spike scales as $\rho(r) \sim r^{-2}$ \cite{Quinlan:1994ed}.
\section{Spike formation for fixed parameter choices}\label{app:formation}
\begin{figure}[t!]
  \centering
\includegraphics[width=0.49\linewidth]{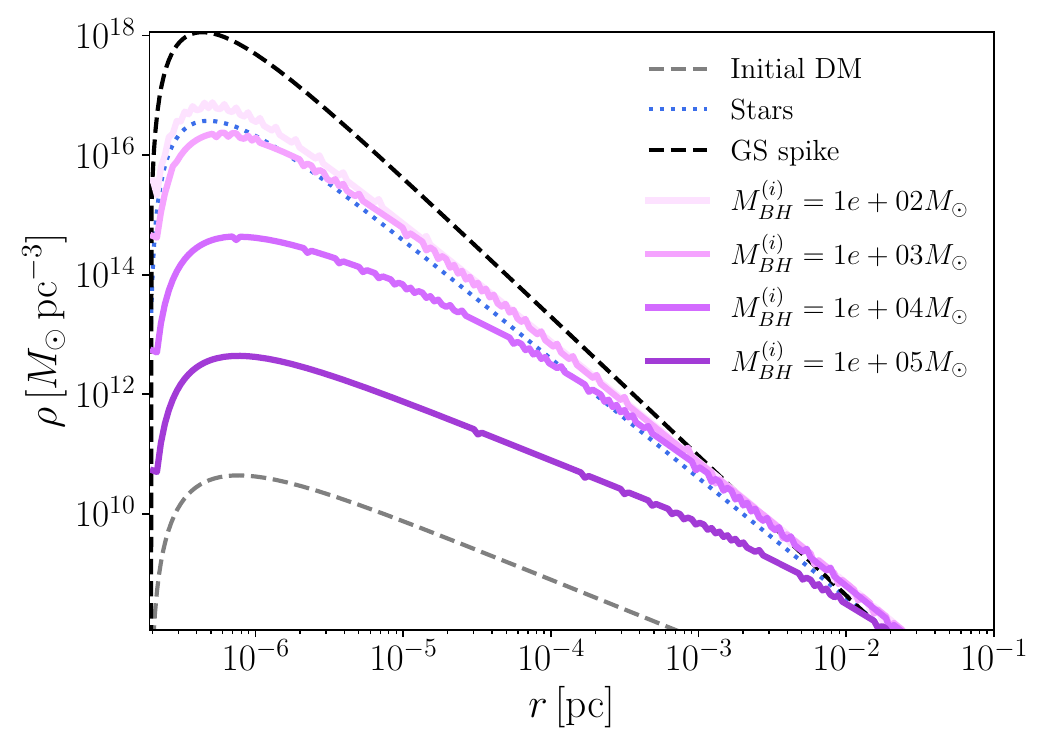} 
\includegraphics[width=0.49\linewidth]{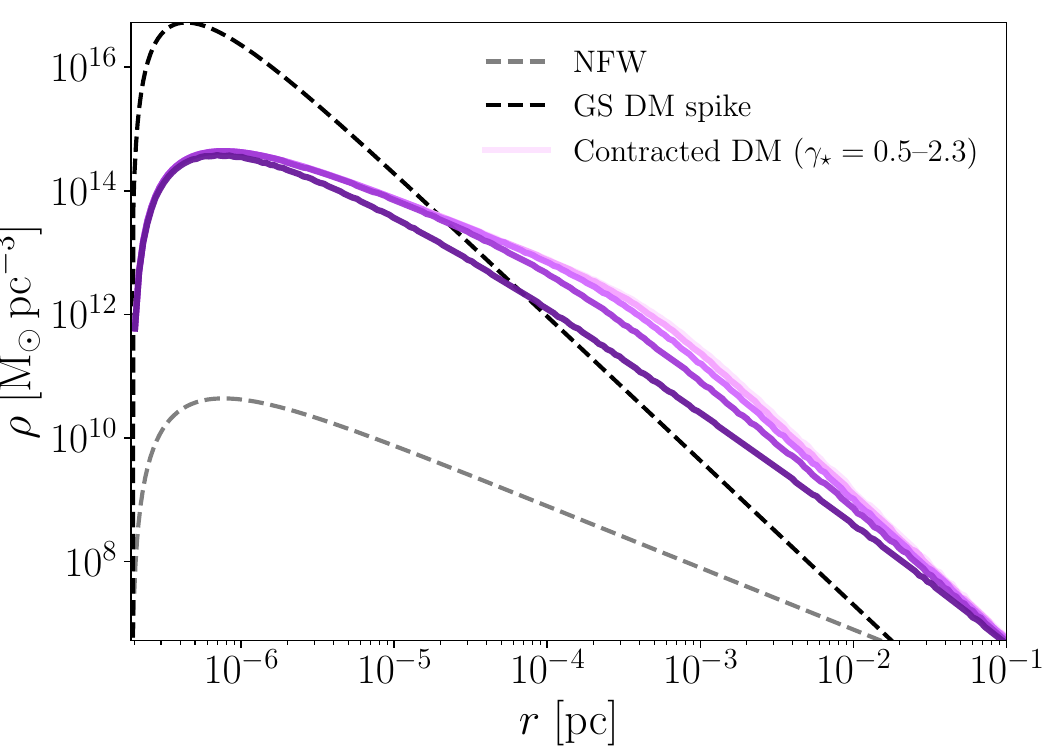} 
\includegraphics[width=0.49\linewidth]{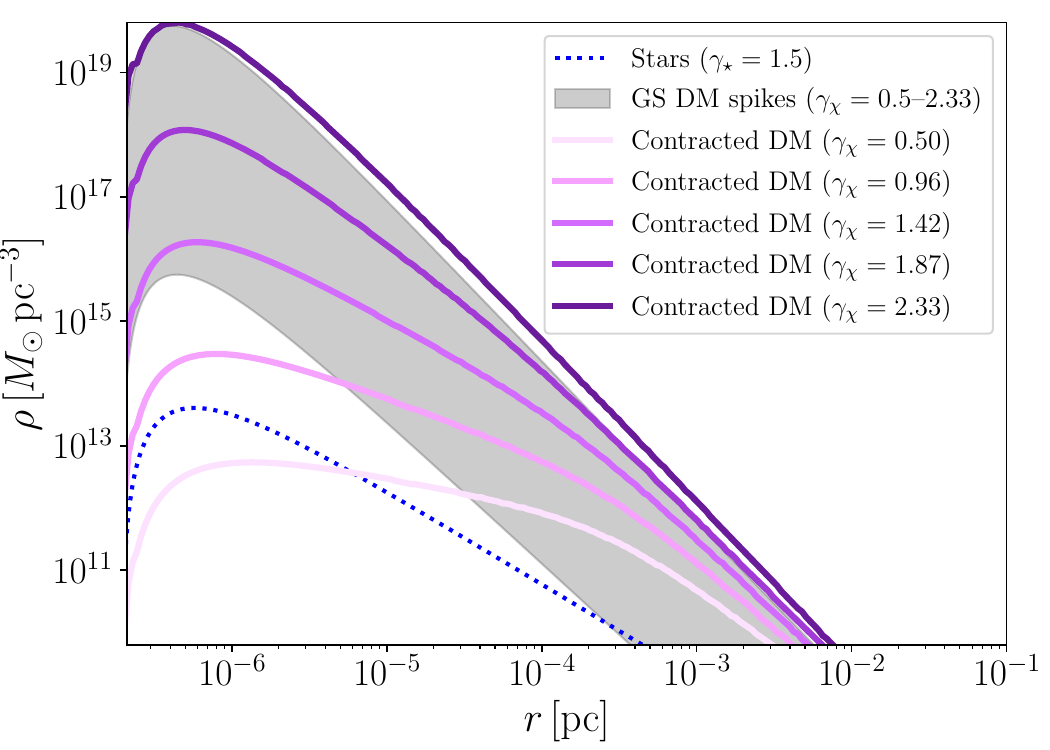} 
\includegraphics[width=0.49\linewidth]{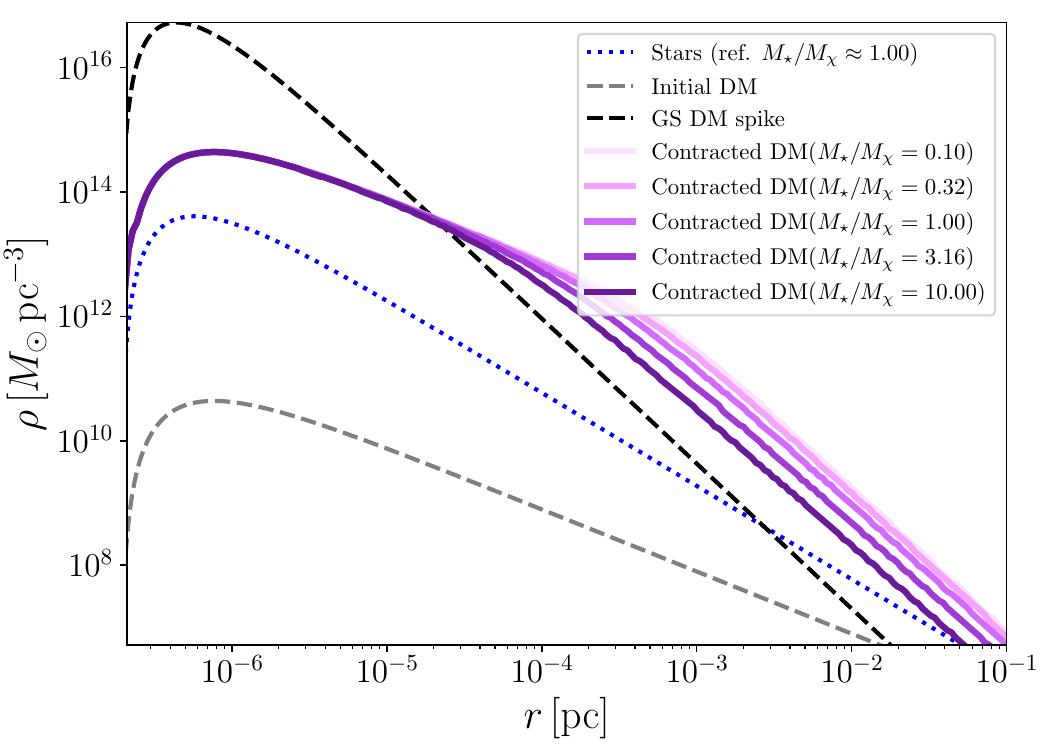} 
  \caption{\textit{Upper left panel:} In purple palette, DM overdensity as a function of distance from the central SMBH for different initial black hole mass seeds. For comparison, we show the initial NFW DM density profile in a dashed grey line, and the initial cuspy profile of stars assumed in a blue dotted lines. The canonical DM spike arising in the Gondolo and Silk prescription is show as a dashed black line. The final black hole mass considered is $10^{6} M_{\odot}$. The normalization of the DM and stellar density profiles is chosen such that their total masses enclosed within 1 pc are equal. \textit{Upper right panel:} Effects of the stellar initial density profile on the final overdensity profile, for a range of values $\gamma_{\star}=0.5-2.3$. The initial BH seed mass taken is $M_{\rm BH}=10^{4}M_{\odot}$. \textit{Lower left panel:} Effects of the initial DM density profile on the final DM overdensities, for a range of values $\gamma_{\chi}=0.5-2.33$. \textit{Lower right panel:}  Effects of the relative normalizations of the DM and stellar density profiles in the final DM profiles, for ratios of enclosed masses within 1 pc in the range $M_{\star}/M_{\chi}=0.1-10$.}
  \label{fig:formation}
\end{figure}

\begin{figure*}[t!]
  \centering
\includegraphics[width=0.49\linewidth]{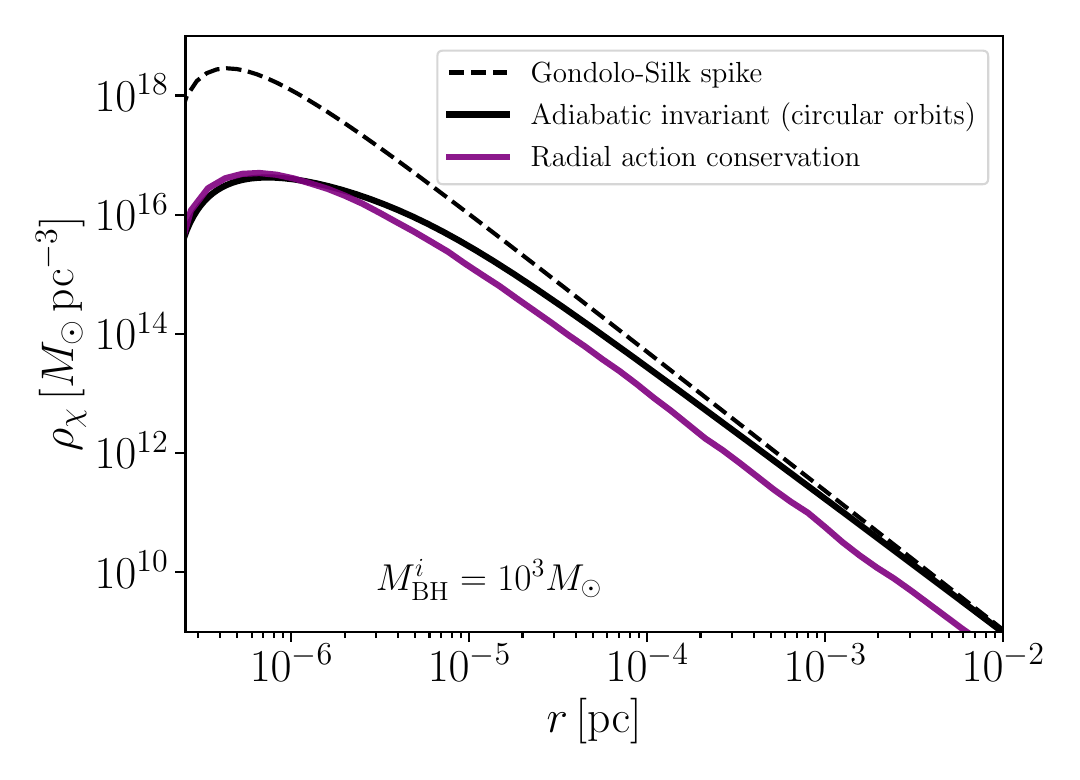}
\includegraphics[width=0.49\linewidth]{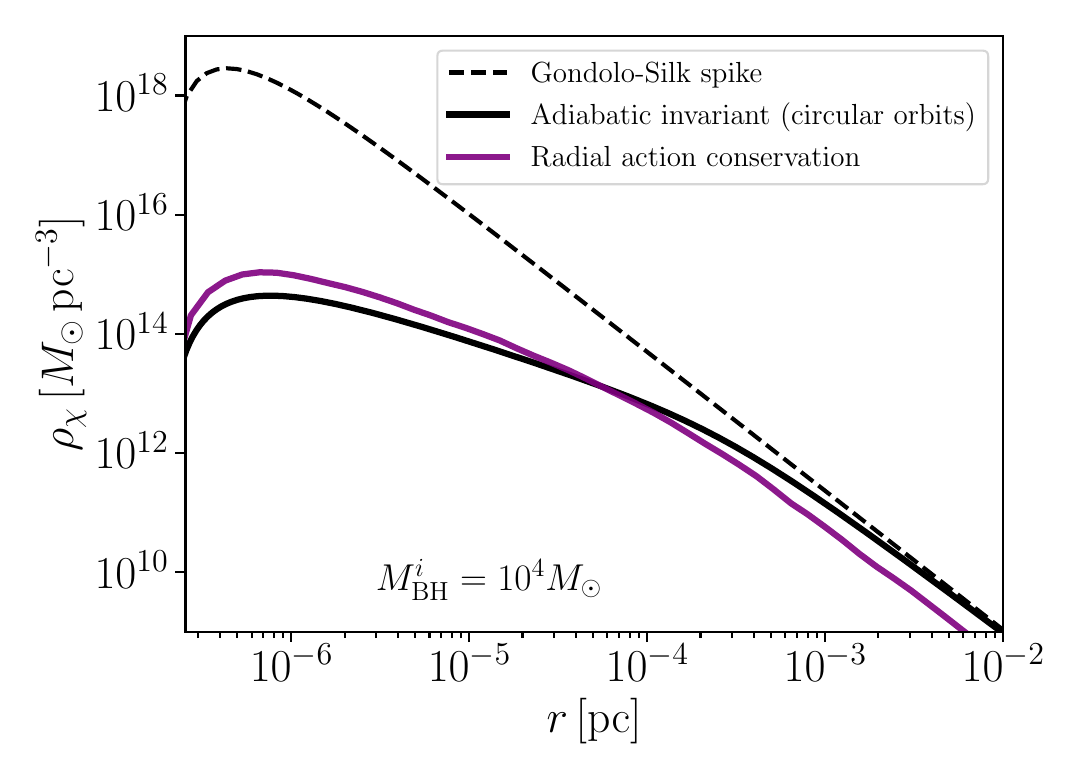}

\caption{DM spikes obtained for three different prescriptions and two choices of initial black hole seed mass. The dashed black line corresponds to the massless black hole seed and DM-only environment from Gondolo and Silk \cite{Gondolo:1999ef}. The solid black lines show the result obtained when considering a massive initial black hole seed and an initial stellar density profile, applying the adiabatic invariant from Eq.~\ref{eq:adiabatic_invariant}, valid for circular orbits. The purple solid lines are numerically obtained from conservation of the radial action for all orbits, where the initial phase-space distribution function prior to black hole growth is obtained by applying Eddington inversion. Both panels consider a final black hole mass of $M_{\rm BH}^{f}=10^{6}M_{\odot}$, an initial stellar density profile with index $\gamma_{\star}=1.5$, DM density profile index of $\gamma_{\chi}=1$, and equal enclosed masses at $r=1$\,pc.}

  \label{fig:full_radial_action}
\end{figure*}

In the main text we derived averaged properties of DM spikes from marginalizing over the uncertain initial conditions of the system. Here we isolate each of these uncertainties to gain a better understanding of their impact on the resulting DM spikes. In Fig.~\ref{fig:formation} we restrict ourselves to the circular-orbit adiabatic invariant and vary one parameter at a time. In Fig.~\ref{fig:full_radial_action} we compare the circular-orbit and full radial-action prescriptions for two representative seed masses.

In the upper left panel of the Figure, we vary the initial BH seed mass for fixed benchmark choices of the remaining parameters. As expected, we find that the lighter the initial seed, the closer the resulting DM overdensity resembles a spike. For heavier seeds, however, the change in the central gravitational potential is smaller, resulting in a less dense DM profile. Our results are qualitatively comparable to those from Ref. \cite{Bertone:2024wbn}, where an initial non-zero black hole mass seed was considered in the formation of the DM overdensities, though without accounting for a surrounding stellar profile as ours. 
In the upper right panel of the Figure, we show the effects of varying the initial stellar profile on the final resulting DM profiles. We notice that the stellar profile index changes at intermediate radii. Near the radius of gravitational influence of the black hole, as expected, all profiles converge, since at that radii the density is controlled by the total enclosed mass, which is the same for all profiles at that radii. At small radii, in the innermost vicinity of the black hole, the profiles also converge. This is expected since in that region the potential is dominated by the central black hole, which is the same for all cases.
In the lower left panel of the Figure, we show the impact of the initial DM density steepness on the final profiles, and confront them with the canonical spike expectations for these cases, shown as a grey band. As expected, the shallower the initial DM profile begins, the less steep it becomes after adiabatic growth. The difference between the GS prescription and our prescription becomes more prominent for shallow initial DM profiles, for which we find that the resulting density profiles are significantly less cuspy than expected in the absence of stars and a zero-mass seed. 
In the lower right panel of the Figure, we show the dependence of the final profiles with the total ratio of enclosed stellar to DM masses within 1 pc. The enclosed total mass choice changed the normalization of the density profiles at large radii, where the potential is not solely dominated by the black hole, and stars also play a role. The larger the enclosed stellar mass, the smaller the resulting DM density in the outer radii is.

\section{Derivation of the steady-state solutions}
\label{app:steady_state_solution}

Here we present an analytic derivation of the steady-state solutions of the isotropic Fokker-Planck equation for a stellar system surrounding a supermassive black hole, and a collisionless DM component heated by the stellar background. Inside the black hole region of influence, the potential is Keplerian,
\begin{equation}
\Phi(r) = -\frac{G M_{\rm BH}}{r}.
\end{equation}
The isotropic density of states and its cumulative are \cite{2008gady.book.....B, Baes:2004wb}
\begin{align}
p(E) &= \frac{\pi}{2\sqrt{2}} (G M_{\rm BH})^3\, E^{-5/2},\\
q(E) &= \frac{\pi}{3\sqrt{2}} (G M_{\rm BH})^3\, E^{-3/2},
\end{align}
In the following, we define the constants $C_p=\frac{\pi}{2\sqrt{2}} (G M_{\rm BH})^3$ and $C_q=\frac{\pi}{3\sqrt{2}} (G M_{\rm BH})^3$. The isotropic, orbit-averaged Fokker--Planck equation in energy space is
\begin{equation}
4\pi^2 p(E)\frac{\partial f}{\partial t}
= -\frac{\partial F_E}{\partial E},
\quad
F_E = -D_{EE}(E)\frac{\partial f}{\partial E} + D_E(E) f.
\end{equation}

For a single stellar mass $M_\star$, the energy diffusion and drift coefficients are \cite{1976ApJ...209..214B,2008gady.book.....B}
\begin{align}
D_{EE}^\star(E) &= 
64\pi^4 G^2 M_\star^2 \ln\Lambda
\left[
q(E)\int_0^E f_\star(E')\,dE' +
\int_E^\infty q(E') f_\star(E')\,dE'
\right],
\\
D_{E}^\star(E) &=
64\pi^4 G^2 M_\star^2 \ln\Lambda
\int_0^E p(E') f_\star(E')\,dE',
\end{align}
where $\ln\Lambda \simeq 15$ is the Coulomb logarithm. A steady-state is then defined by a vanishing energy flux, i.e
\begin{equation}
F_E = 0.
\end{equation}

\subsection{Bahcall-Wolff solution}

We begin by deriving the steady-state solution for stars. We assume a power-law form for the stellar distribution function,
\begin{equation}
f_\star(E) = A E^{k_\star}.
\end{equation}
The relevant integrals are
\begin{align}
\int_0^E f_\star(E')\,dE' &= \frac{A}{k_\star+1} E^{k_\star+1}, \\[6pt]
\int_E^\infty q(E') f_\star(E')\,dE' &= 
\frac{A C_q}{\tfrac{1}{2}-k_\star} E^{k_\star - 1/2},
\quad (k_\star < \tfrac{1}{2}), \\[6pt]
\int_0^E p(E') f_\star(E')\,dE' &=
\frac{A C_p}{k_\star - \tfrac{3}{2}} E^{k_\star - 3/2}.
\end{align}
Substituting into the diffusion coefficients yields
\begin{align}
D_{EE}^\star(E) &= 
\mathcal{C}\, A\, C_q\, E^{k_\star - 1/2}
\left[
\frac{1}{k_\star+1} + \frac{1}{\tfrac{1}{2}-k_\star}
\right], \\[6pt]
D_E^\star(E) &=
-\mathcal{C}\, A\, C_p\,
\frac{E^{k_\star - 3/2}}{k_\star - \tfrac{3}{2}}.
\end{align}
The steady-state condition $F_E=0$ gives
\begin{equation}
D_{EE}^\star \frac{\partial f_\star}{\partial E}
+ D_E^\star f_\star = 0,
\end{equation}
and, using $\partial_E f_\star = A k_\star E^{k_\star-1}$, and after some algebra, one finds
\begin{equation}
\frac{2k_\star}{3}
\left(
\frac{1}{k_\star+1} + \frac{1}{\tfrac{1}{2}-k_\star}
\right)
+ \frac{1}{k_\star - \tfrac{3}{2}} = 0.
\end{equation}
The unique physical solution with $k_\star < 1/2$ is
\begin{equation}
k_\star = \frac{1}{4}.
\end{equation}

For an isotropic distribution in a Kepler potential,
\begin{equation}
\rho_\star(r) \propto r^{-(k_\star + 3/2)},
\end{equation}
so that
\begin{equation}
\rho_\star(r) \propto r^{-7/4}.
\end{equation}

\subsection{Gnedin-Primack-Merritt solution}

The energy diffusion of DM particles is driven by stars, hence
\begin{equation}
D_{EE}^\chi(E) = D_{EE}^\star(E).
\end{equation}
However, to leading order in $m_\chi/m_\star$, the dynamical friction (drift) term vanishes (see Appendix~\ref{sec:diffusion_coefficients}),
\begin{equation}
D_E^\chi(E) \simeq 0.
\end{equation}

The DM energy flux therefore reads
\begin{equation}
F_E^\chi = -D_{EE}^\star(E)\frac{\partial f_\chi}{\partial E}.
\end{equation}
From the steady-state condition $F_E^\chi = 0$, one gets
\begin{equation}
D_{EE}^\star(E)\frac{\partial f_\chi}{\partial E} = 0 \Rightarrow \frac{\partial f_\chi}{\partial E} = 0,
\end{equation}
since $D_{EE}^\star(E)\neq 0$ inside the cusp. Then
\begin{equation}
f_\chi(E) = \text{const.} \quad \Rightarrow \quad k_\chi = 0.
\end{equation}
Using the Keplerian mapping between phase space and real space,
\begin{equation}
\rho(r) \propto r^{-(k + 3/2)},
\end{equation}
we obtain
\begin{equation}
\rho_\chi(r) \propto r^{-3/2}.
\end{equation}

\section{Redshift evolution of density profiles for fixed parameters}\label{app:fixed_evolution}

\begin{figure}[t!]
  \centering
\includegraphics[width=0.49\linewidth]{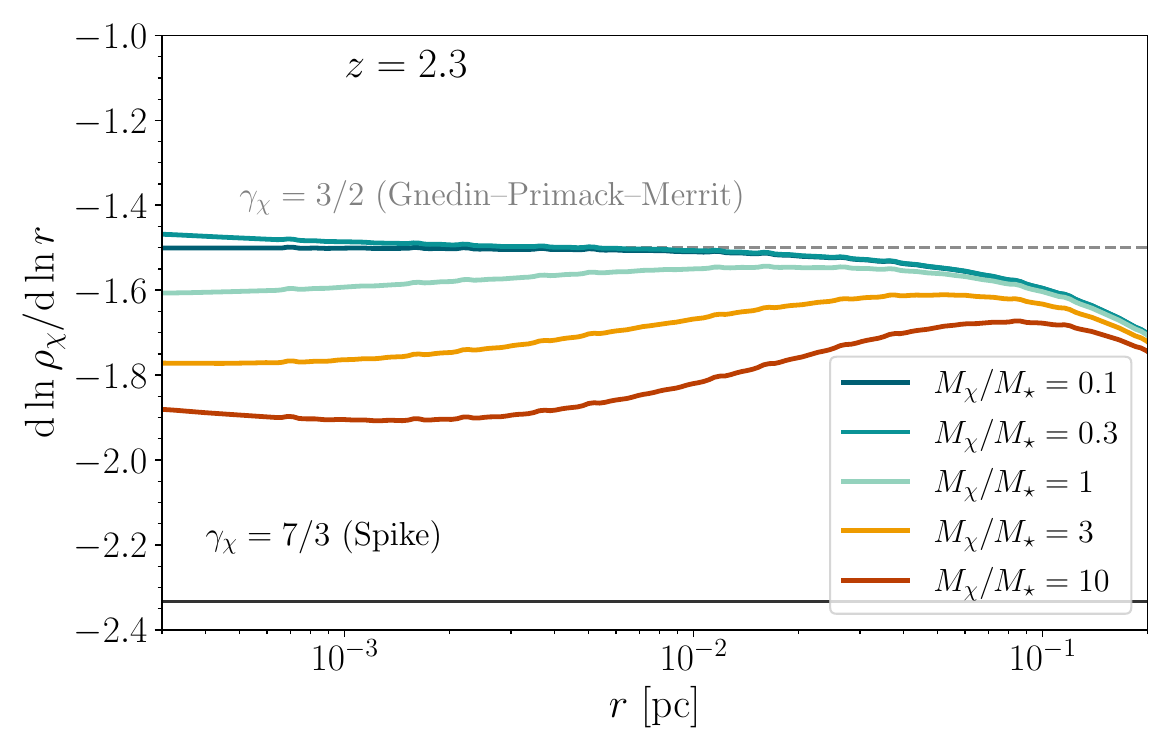}
\includegraphics[width=0.49\linewidth]{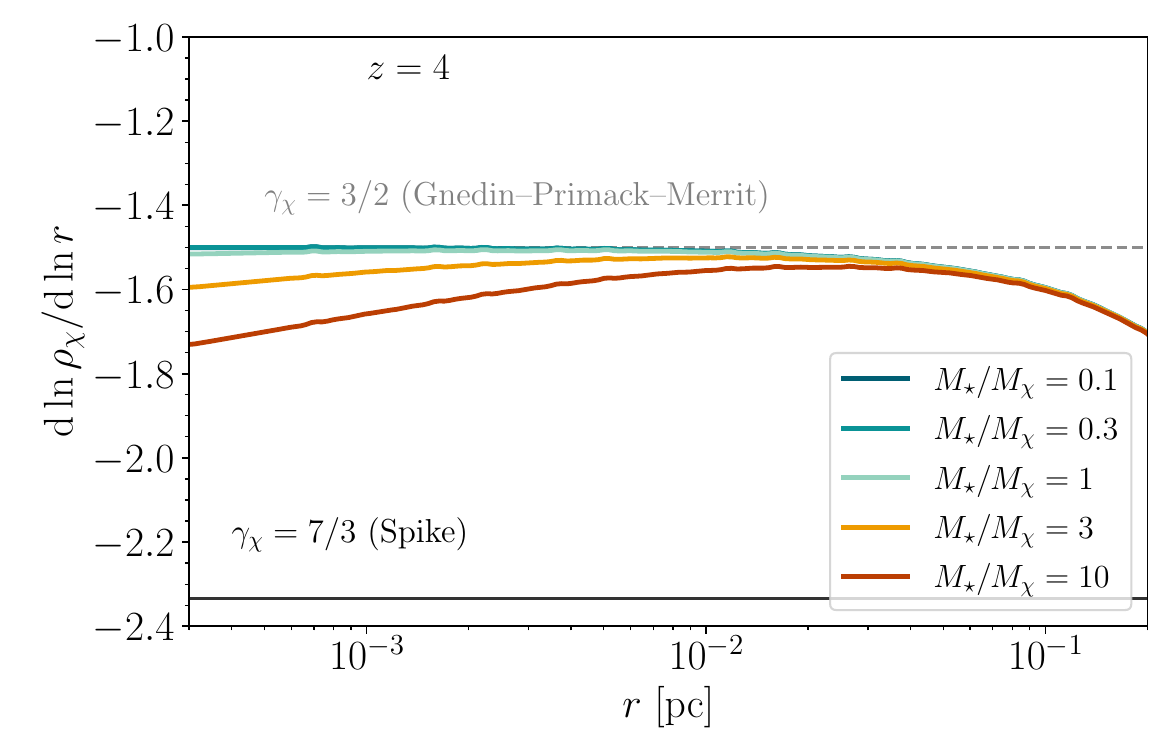}
\includegraphics[width=0.49\linewidth]{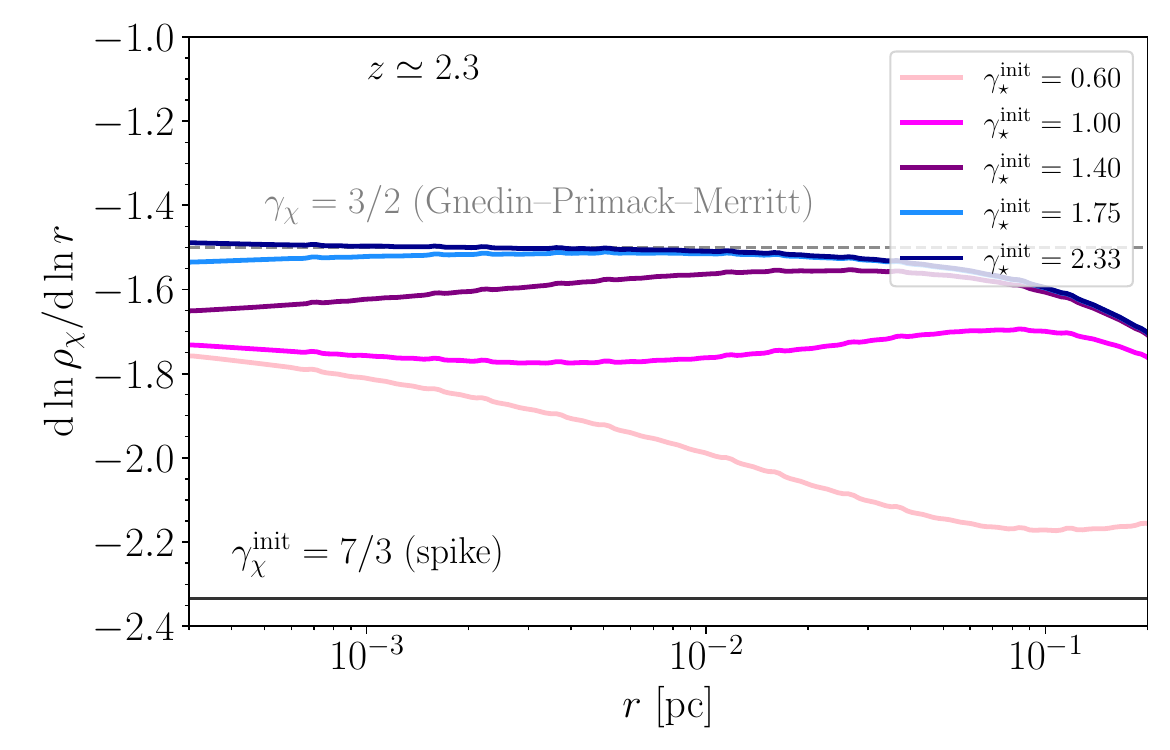}
\includegraphics[width=0.49\linewidth]{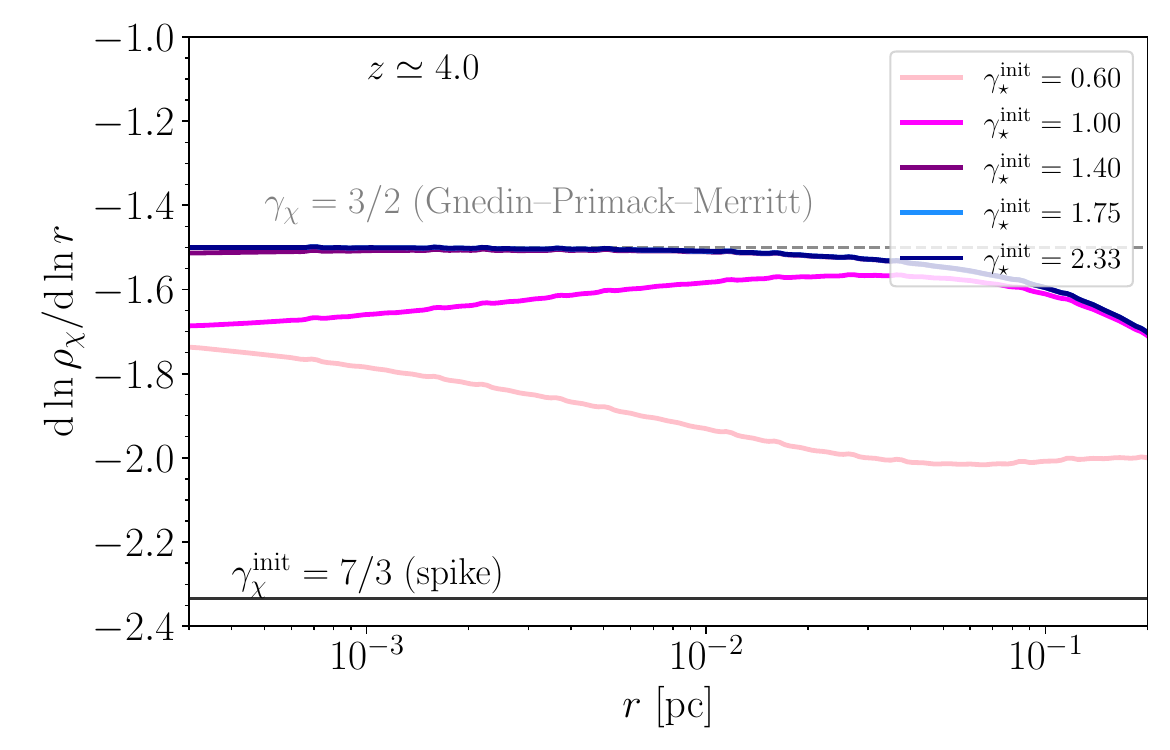}
  \caption{\textit{Left panel:} DM density profile inner slope index at different distances from the central black hole, for different ratios of enclosed stellar to DM masses in the region of gravitational influence of the black hole, and for a system initialized at $z=10$. We show the evolved profiles at timestamp $z=2.3$. \textit{Right panel:} Same as left panel, but tracking the stellar normalization across redshift with the SFR, and fixing the redshift in this plot at $z=4$.}
  \label{fig:inner_slope_mass_ratio_app}
\end{figure}

In Fig.~\ref{fig:inner_slope_mass_ratio_app} we quantify how the dark-matter density profile index responds to stellar heating as a function of radius, for different choices of initial conditions. We define the logarithmic slope,
\begin{equation}
\gamma_\chi(r) \equiv -\frac{{\rm d}\ln\rho_\chi}{{\rm d}\ln r},
\end{equation}
evaluated from the evolved profiles at fixed redshift after an initial spike with $\gamma_{\rm sp}=7/3$ has formed at $z=10$. The upper left panel assumes a continuous stellar bath with fixed normalization and compares several values of the enclosed stellar-to-DM mass ratio within the radius of influence, $M_\star/M_\chi$. Larger stellar masses produce more efficient heating, leading to a faster convergence towards the steady-state index $\gamma_\chi\simeq 3/2$ at small radii, while DM-dominated configurations preserve steeper slopes ($\gamma_\chi \gtrsim 2$) down to smaller scales. At larger radii, $r\gtrsim r_h$, all models asymptote back to the original halo slope, reflecting the fact that diffusion is inefficient where the stellar density is low. The upper right panel shows the same quantity when the heating rate is modulated by the cosmic SFR, evaluated at a different fixed redshift. In this case, the cusp relaxes more rapidly around the SFR peak ($z\simeq2$--3), and the transition between the relaxed inner region and the unperturbed outer halo becomes sharper. This demonstrates that the final radial slope is strongly dependent on the relative enclosed stellar mass. The lower left panel fixes the ratio of enclosed stellar to DM masses to one, but varies the initial stellar profile. We note that shallower stellar profiles lead to a milder relaxation of the inner profiles, still preserving partially the spike around redshift $z \simeq 2.3$. However, initially steep stellar profiles $\gamma_{\star} \gtrsim 1.5$ lead the DM profiles to reach the steady state solution by redshift $z \simeq 2.3$. In the lower right panel this is shown for a redshift dependent stellar bath, and it can be noticed that by redshift $z \simeq 4$ the steady state solution has already been reached for most initial stellar profiles.

\begin{figure}[t!]
  \centering
\includegraphics[width=0.49\linewidth]{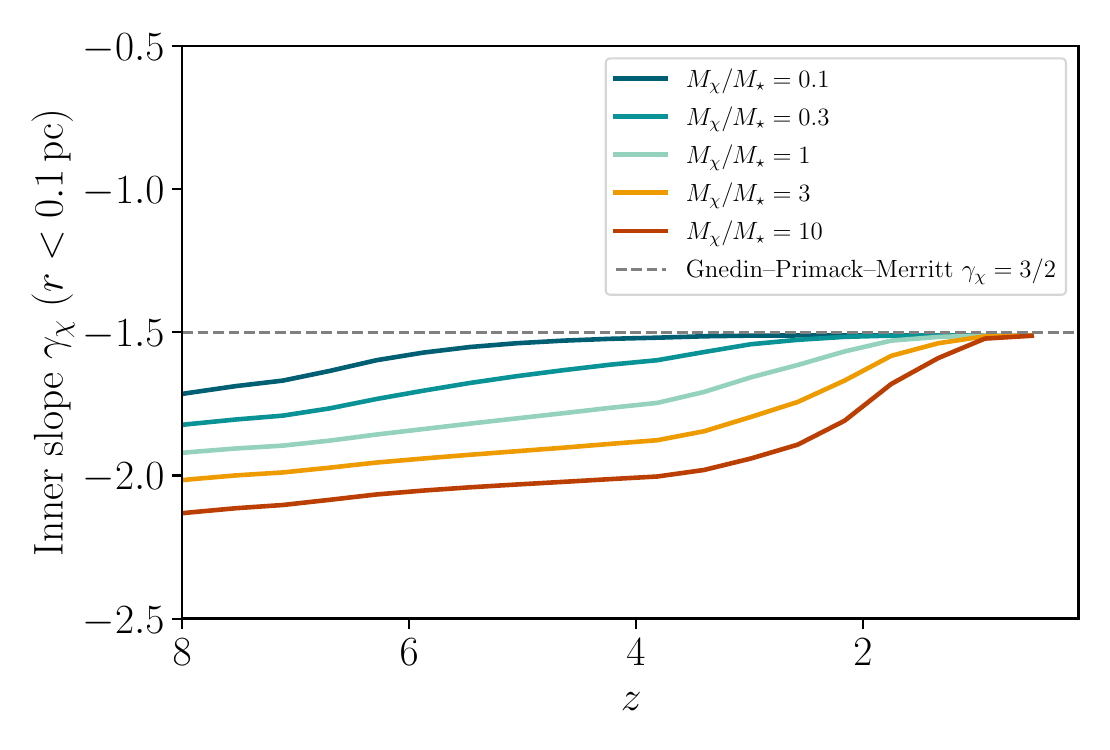} 
\includegraphics[width=0.49\linewidth]{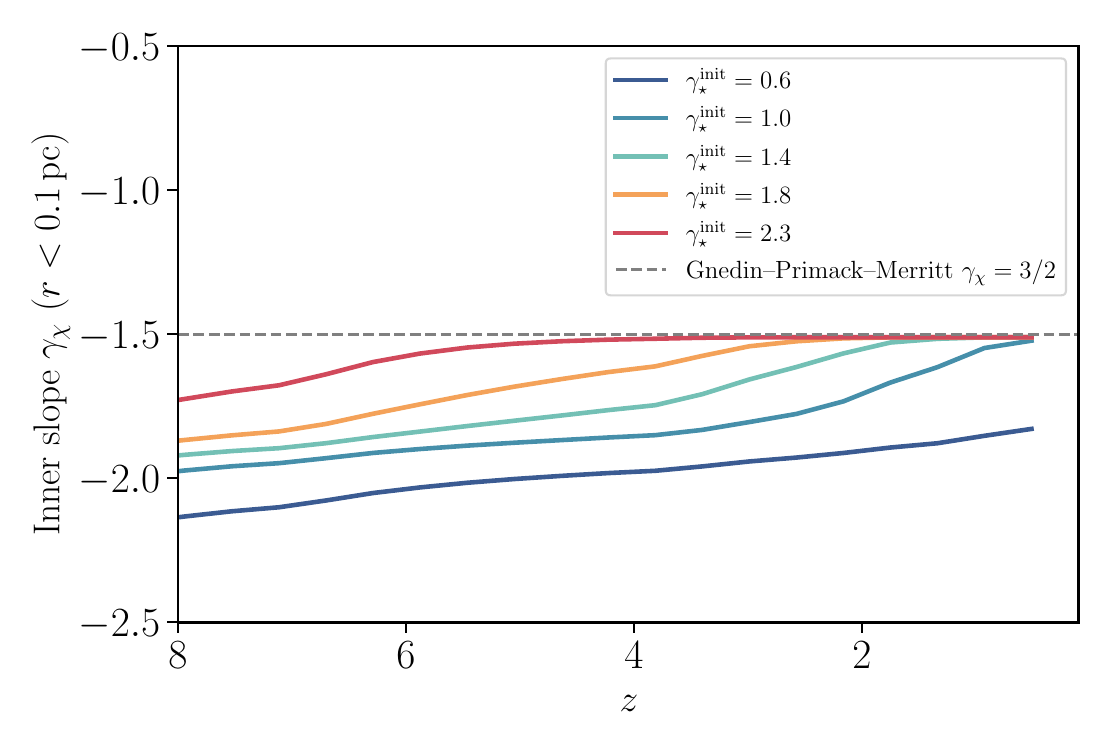} 
\includegraphics[width=0.49\linewidth]{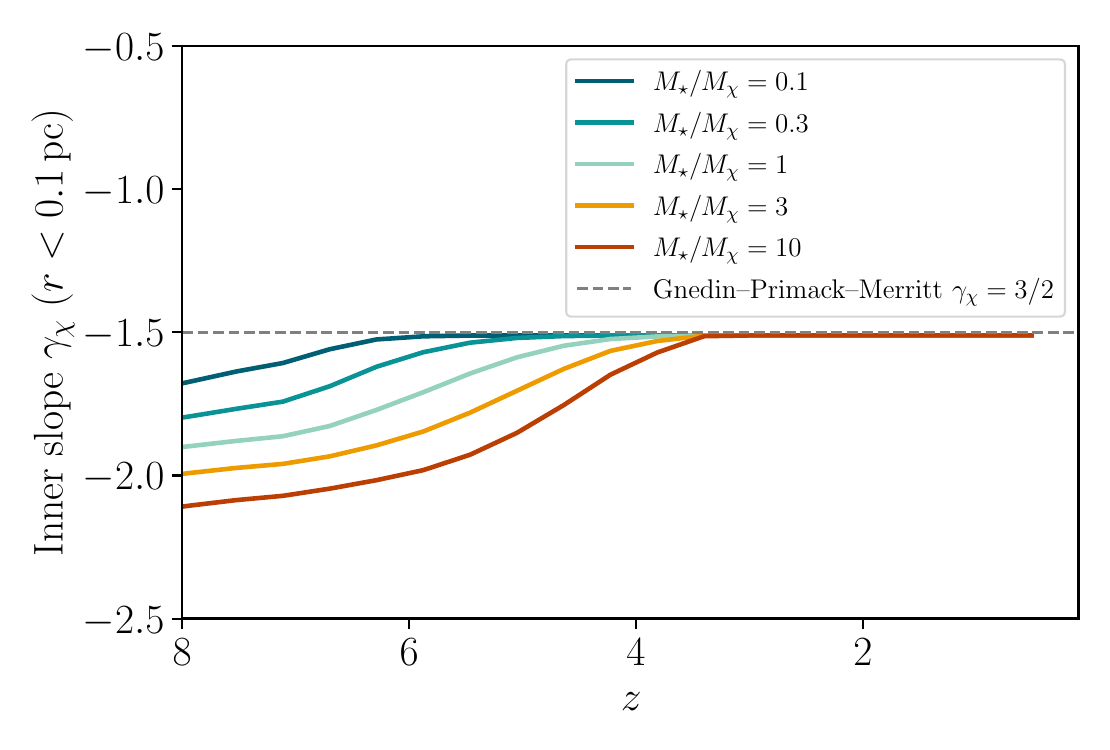} 
\includegraphics[width=0.49\linewidth]{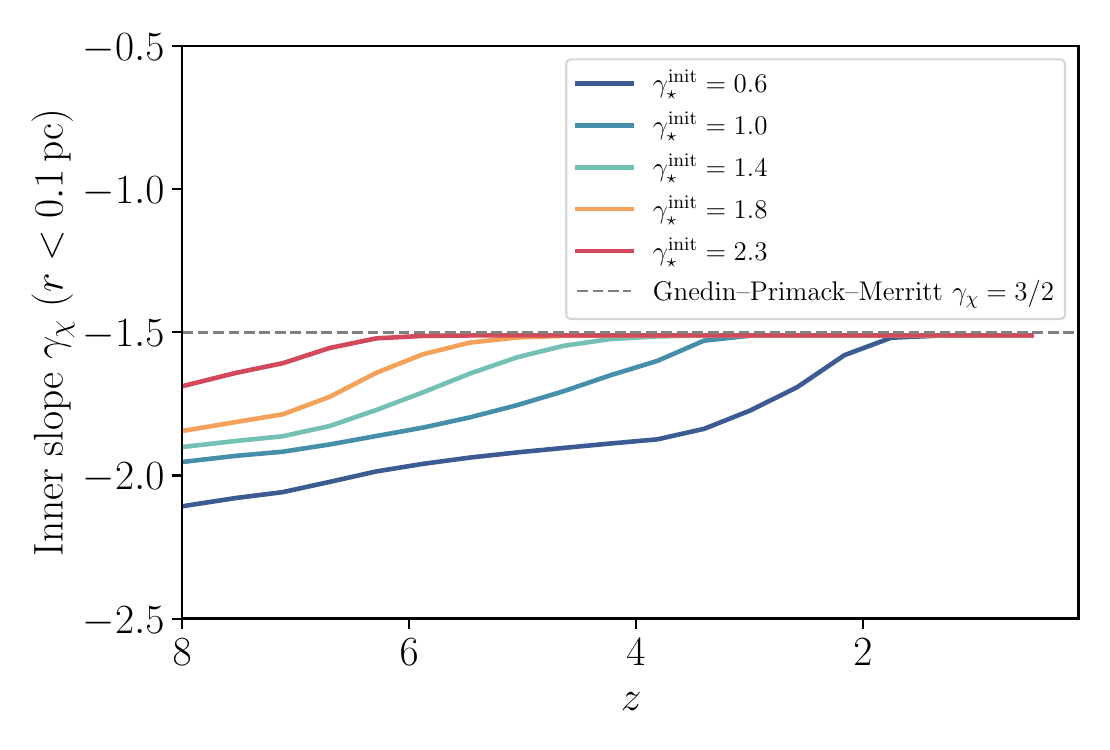} 
  \caption{\textit{Upper left panel:} Evolution of the inner slope DM profile index within $r<0.1$pc as a function of redshift, for different values of the enclosed DM to stellar masses within the region of gravitational influence of the black hole, and a constant stellar heating bath. For comparison, we also show the steady state solution at $\gamma=3/2$ as a dashed grey line. \textit{Upper right panel:} Same as the right panel, but fixing the enclosed mass ratio to $M_{\chi}/M_{\star}=1$, and varying the stellar initial index in the range $\gamma_{\star}=0.6-2.3$. \textit{Lower panels:} Same as the right panels, but considering a stellar heating rate weighted over redshift by the SFR.}
  \label{fig:inner_slope_redshift_app}
\end{figure}

Figure~\ref{fig:inner_slope_redshift_app} summarizes the global evolution of the inner cusp by showing the volume-averaged logarithmic slope $\langle\gamma_\chi\rangle$ within $r<0.1~{\rm pc}$ as a function of redshift for a range of model parameters. The upper panels correspond to a continuous stellar bath with fixed normalization: the left panel varies the enclosed mass ratio $M_\chi/M_\star$ at fixed stellar index $\gamma_\star$, while the right panel varies $\gamma_\star$ at fixed $M_\chi/M_\star=1$. In both cases, systems with more massive or cuspier stellar components relax more quickly, reaching $\langle\gamma_\chi\rangle\simeq 1.5$ already by $z\sim 2$--3, whereas DM-dominated or shallow stellar profiles retain steeper cusps for a longer time. The lower panels repeat the exercise when the diffusion rate is weighted by the cosmic SFR. In this scenario, the relaxation accelerates around the SFR peak and slows down at early and late times, but the qualitative behavior is unchanged: independently of the precise choice of $M_\chi/M_\star$ and $\gamma_\star$, the inner slope converges towards the Bahcall--Wolf value $\langle\gamma_\chi\rangle\simeq 3/2$ by $z\lesssim 2$, although for most scenarios the steady state solution is reached even earlier, by redshift $z \simeq 2$. We also notice that, the larger the enclosed stellar mass or the steepness of the stellar profile, the more likely it is that the initial DM spike is quickly relaxed to a core, which then slowly regrows towards the relaxation solution. However, in scenarios where the stellar component is underdense or shallow, the relaxation of the DM spike is continuous and smooth towards the steady state solution.

\section{Derivation of diffusion coefficients in energy space}\label{sec:diffusion_coefficients}

In this Appendix we derive the diffusion coefficients appearing in the
orbit-averaged Fokker-Planck equation used in the main text, and clarify
their relation to the velocity-space diffusion coefficients.
Our goal is to justify the neglect of the first--order (drift) term in the
DM Fokker-Planck equation when DM is heated by a stellar
background.

Throughout this Appendix we work directly with the specific orbital
energy,
\begin{equation}
E = \frac{v^2}{2} + \Phi(\mathbf{x}),
\end{equation}
which is the variable employed in the main text (with sign conventions
such that $E<0$ for bound orbits).

We begin by recalling the standard velocity--space diffusion coefficients
for a test particle of mass $m$ moving through a background of field
particles of mass $m_a$. Denoting by $\Delta\mathbf v$ the velocity
increment accumulated over a short time interval
\cite{2008gady.book.....B}
\begin{align}
D[\Delta v_{\parallel}]
&=
-\frac{16 \pi^2 G^2 m_a (m+m_a)\ln\Lambda}{v^2}
\int_0^v \mathrm{d}v_a\, v_a^2 f_a(v_a),
\label{eq:Dvpar_app}
\\
D\!\left[(\Delta v_{\parallel})^2\right]
&=
\frac{32 \pi^2 G^2 m_a^2\ln\Lambda}{3}
\left[
\int_0^v \mathrm{d}v_a\,\frac{v_a^4}{v^3} f_a(v_a)
+
\int_v^\infty \mathrm{d}v_a\, v_a f_a(v_a)
\right],
\label{eq:Dvpar2_app}
\\
D\!\left[(\Delta\mathbf v_\perp)^2\right]
&=
\frac{32 \pi^2 G^2 m_a^2\ln\Lambda}{3}
\left[
\int_0^v \mathrm{d}v_a
\left(\frac{3v_a^2}{v}-\frac{v_a^4}{v^3}\right) f_a(v_a)
+
2\int_v^\infty \mathrm{d}v_a\, v_a f_a(v_a)
\right].
\label{eq:Dvperp2_app}
\end{align}
The first coefficient describes dynamical friction, while the second--order
coefficients encode velocity diffusion. Importantly, these terms exhibit
different dependences on the test--particle mass.

Since gravitational encounters induce stochastic changes in the velocity,
the corresponding increment in specific energy must be expanded to second
order in $\Delta\mathbf v$,
\begin{equation}
\Delta E
=
\mathbf v\cdot\Delta\mathbf v
+
\frac{1}{2}(\Delta\mathbf v)^2 .
\label{eq:deltaeps}
\end{equation}
The first-- and second--order diffusion coefficients in specific energy
space are defined as
\begin{equation}
D_E \equiv \frac{\langle\Delta E\rangle}{\Delta t},
\qquad
D_{EE} \equiv
\frac{\langle(\Delta E)^2\rangle}{2\,\Delta t}.
\end{equation}

The first moment can then be written as
\begin{equation}
D_E
=
v\,D[\Delta v_\parallel]
+
\frac{1}{2}D[(\Delta\mathbf v)^2],
\label{eq:Deps_full}
\end{equation}
with
\begin{equation}
D[(\Delta\mathbf v)^2]
=
D[(\Delta v_\parallel)^2]
+
D[(\Delta\mathbf v_\perp)^2].
\end{equation}

Adding Eqs.~\eqref{eq:Dvpar2_app} and \eqref{eq:Dvperp2_app}, the terms
proportional to $v_a^4/v^3$ cancel exactly, yielding
\begin{equation}
D[(\Delta\mathbf v)^2]
=
32\pi^2 G^2 m_a^2\ln\Lambda
\left[
\frac{1}{v}\int_0^v \mathrm{d}v_a\, v_a^2 f_a(v_a)
+
\int_v^\infty \mathrm{d}v_a\, v_a f_a(v_a)
\right].
\label{eq:Dv2_app}
\end{equation}

Substituting Eqs.~\eqref{eq:Dvpar_app} and \eqref{eq:Dv2_app} into
Eq.~\eqref{eq:Deps_full}, one finds
\begin{align}
D_E
&=
-\frac{16\pi^2 G^2 m_a (m+m_a)\ln\Lambda}{v}
\int_0^v \mathrm{d}v_a\, v_a^2 f_a(v_a)
\nonumber\\
&\quad
+
16\pi^2 G^2 m_a^2\ln\Lambda
\left[
\frac{1}{v}\int_0^v \mathrm{d}v_a\, v_a^2 f_a(v_a)
+
\int_v^\infty \mathrm{d}v_a\, v_a f_a(v_a)
\right].
\end{align}
The contributions proportional to $m_a^2$ in the first integral cancel
exactly between the two terms, leaving
\begin{equation}
D_E
=
16\pi^2 G^2 m_a\ln\Lambda
\left[
m_a\int_v^\infty \mathrm{d}v_a\, v_a f_a(v_a)
-
\frac{m}{v}\int_0^v \mathrm{d}v_a\, v_a^2 f_a(v_a)
\right].
\label{eq:Deps_final}
\end{equation}
Note that $D_E$ in this Kramers--Moyal form still contains two
contributions: a term proportional to $m_a^2$ from fast particles in
the bath (first integral) and a dynamical-friction term proportional
to $m\,m_a$ (second integral).  The suppression of the drift
relative to diffusion only becomes apparent once the equation is
recast in conservative form, as we now show.

We now derive the second--order diffusion coefficient. From
Eq.~\eqref{eq:deltaeps},
\begin{align}
(\Delta E)^2
&=
(\mathbf v\cdot\Delta\mathbf v)^2
+ (\mathbf v\cdot\Delta\mathbf v)(\Delta\mathbf v)^2
+ \frac{1}{4}(\Delta\mathbf v)^4 .
\end{align}
Since $\Delta\mathbf v$ is small, the last two terms are higher order and
can be neglected. Keeping only the leading contribution yields
\begin{equation}
(\Delta E)^2 \simeq (\mathbf v\cdot\Delta\mathbf v)^2
= v^2 (\Delta v_\parallel)^2 .
\label{eq:deltaeps2}
\end{equation}

The second--order diffusion coefficient is therefore
\begin{equation}
D_{EE}
\simeq
\frac{v^2}{2}\,
D\!\left[(\Delta v_\parallel)^2\right].
\label{eq:Dee_final}
\end{equation}
Substituting Eq.~\eqref{eq:Dvpar2_app}, one finds
\begin{equation}
D_{EE}
\propto
m_a^2 ,
\end{equation}
independent of the test--particle mass.

The orbit--averaged Fokker--Planck equation can be written in
Kramers--Moyal form,
\begin{equation}
4\pi^2 p(E)\frac{\partial f}{\partial t}
=
-\frac{\partial}{\partial E}\!\left[D_E \, f\right]
+\frac{\partial^2}{\partial E^2}\!\left[D_{EE} \, f\right],
\end{equation}
where $D_E = \langle\Delta E\rangle/\Delta t$ and
$D_{EE} = \langle(\Delta E)^2\rangle/(2\,\Delta t)$.
Expanding the second derivative and collecting terms, this is
equivalently written in conservative (flux) form,
\begin{equation}
4\pi^2 p(E)\frac{\partial f}{\partial t}
=
-\frac{\partial}{\partial E}
\left[
- D_{EE}\,\frac{\partial f}{\partial E}
- \widetilde{D}_E \, f
\right],
\end{equation}
with $\widetilde{D}_E \equiv D_E - \partial_E D_{EE}$. $D_E$ contains a term
proportional to $m_a^2$ from stochastic energy kicks.
Since $D_{EE}\propto m_a^2$, its energy derivative
$\partial_E D_{EE}$ cancels this contribution exactly, so that
$\widetilde{D}_E$ retains only the dynamical-friction piece
proportional to $m\,m_a$.
In the main text we denote $\widetilde{D}_E$ simply as $D_E$;
its ratio to the diffusion coefficient scales as
\begin{equation}
\frac{\widetilde{D}_E}{D_{EE}}
\sim \frac{m}{m_a}
\end{equation}

For DM particles interacting gravitationally with a stellar
background, $m_\chi \ll m_\star$, implying
\begin{equation}
\frac{D_E}{D_{EE}}
\sim
\frac{m_\chi}{m_\star}
\ll 1 .
\end{equation}
The first-order (drift) term in the DM Fokker-Planck equation is
therefore parametrically suppressed and can be safely neglected, while the
second-order diffusion term is finite and entirely driven by the stellar
bath. This justifies the treatment adopted in the main text.

\section{Alternative implementations of Star Forming Rate effects}

The evolution of the DM distribution function $f_\chi(E,t)$ in a
Keplerian potential is governed by
\begin{equation}
\frac{\partial f_\chi}{\partial t}
=
\frac{1}{4\pi^2 p(E)}
\frac{\partial}{\partial E}
\left[
D_{EE}(E,t)\,
\frac{\partial f_\chi}{\partial E}
\right],
\end{equation}
where $p(E)$ is the density of states and $D_{EE}$ is the energy–diffusion
coefficient generated by gravitational scattering with stars,
\begin{equation}
D_{EE}(E,t)
=
64\pi^4 G^2 \ln\Lambda
\left[
q(E)\!\int_0^E\! h(E',t)\, dE'
+
\int_E^\infty\! q(E')\,h(E',t)\, dE'
\right],
\end{equation}
with
\begin{equation}
h(E,t)=m_\star f_\star(E,t).
\end{equation}
The stellar distribution therefore determines the relaxation rate that
drives DM evolution. In the main text, the stellar distribution is assumed to remain fixed while the relaxation efficiency is modulated by the cosmic star formation rate density $\psi(z)$.  The effective timestep entering the evolution is rescaled according to
\begin{equation}
\Delta t_{\rm eff}
=
\Delta t\,
\frac{\psi(z)}{\psi(z_{\rm init})},
\end{equation}
which is equivalent to evolving with a relaxation clock
\begin{equation}
\tau_{\rm eff}(t)
=
\int_0^t
\frac{\psi[z(t')]}{\psi(z_{\rm init})}
\frac{dt'}{T_{\rm heat}},
\end{equation}
where $T_{\rm heat}$ is the stellar heating timescale.  In this prescription, the stellar background is unchanged, but the star formation history modifies the rate at which relaxation proceeds.
\begin{figure}[t!]
  \centering
\includegraphics[width=0.49\linewidth]{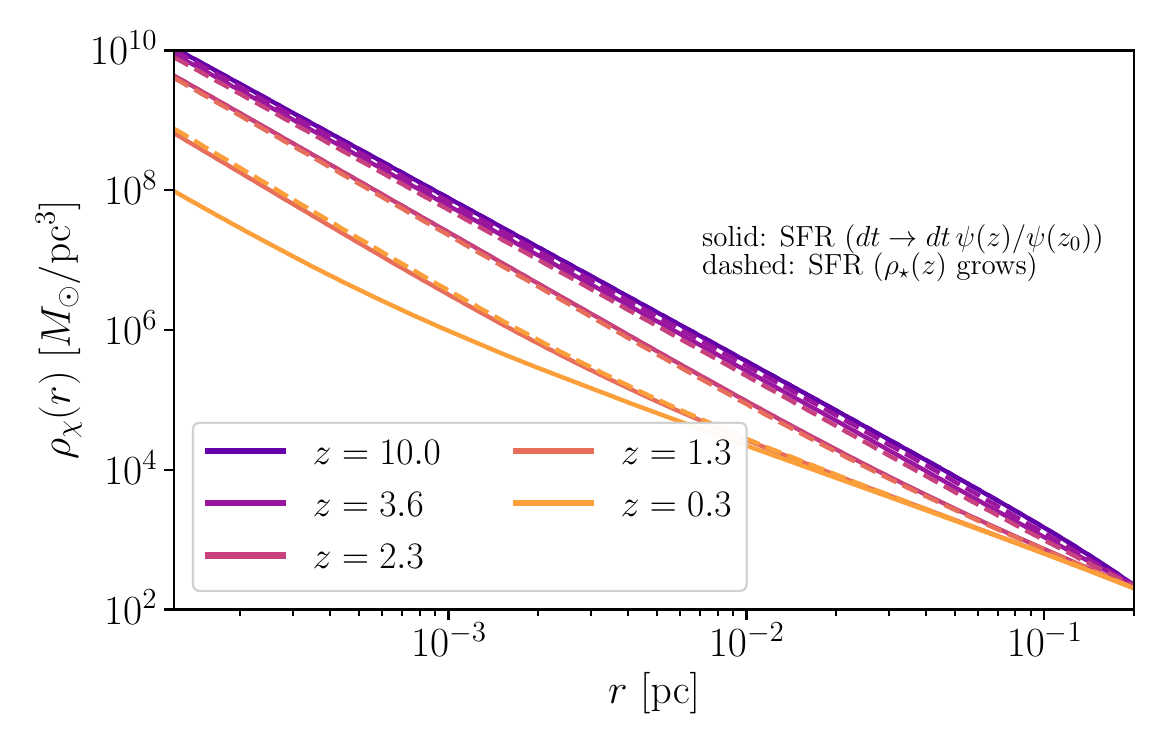}
  \caption{Comparison of different SFR implementations used in this work: the smooth cosmic SFR from Ref.~\cite{Madau:2014bja} and the cumulative stellar mass fraction used to modulate the diffusion coefficient in the alternative normalization prescription.}
  \label{fig:SFRs_implementations}
\end{figure}
Another possible implementation is the following: the star formation history instead controls the build-up of the stellar density itself, which directly modifies the diffusion coefficient through the factor $h(E,t)$.  We implement this approach as well and compare to our previous method, model the stellar density normalization at $1\,{\rm pc}$ as
\begin{equation}
\rho_\star(1{\rm pc},z)
=
\rho_{\star,0}\,
\frac{F_{\rm seed}+f_{\rm formed}(z)}
     {F_{\rm seed}+1},
\end{equation}
where $\rho_{\star,0}$ is the present-day normalization, $F_{\rm seed}$ represents the fraction of stellar mass already present at the initial redshift, and
\begin{equation}
f_{\rm formed}(z)
=
\frac{\int_z^{z_{\rm init}} \psi(z')\,dz'}
     {\int_0^{z_{\rm init}} \psi(z')\,dz'}
\end{equation}
is the cumulative fraction of stellar mass assembled since the initial epoch.  Since $D_{EE}\propto m_\star f_\star \propto \rho_\star$, this approach modifies the diffusion coefficient directly rather than introducing an effective timestep.

The results of this prescription are shown for a set of initial conditions in Fig.~\ref{fig:SFRs_implementations}, where we see that the cumulative-mass implementation relaxes the profiles slower than the prescription discussed in the main text.

\section{Convergence tests}\label{app:convergence}
When solving Eq. \ref{eq:FP_master} and Eq. \ref{eq:FP_star}, we introduce numerical uncertainties through some discretization and solver parameters. First, the Fokker-Planck equations are discretized on a logarithmic grid in orbital energy $E$, with $N_E$ bins. Throughout the main text, we choose $N_E=480$. This parameter controls the resolution in phase space. Second, the time evolution of the phase-space distribution is divided into $n_{\rm steps}$ timesteps, which fixes the timestep size $\Delta t$ in units of the stellar heating timescale as defined in Eq. \ref{eq:tau_def}, and which we convert in redshift after assuming an initial redshift corresponding to $\tau_0$. Increasing $n_{\rm steps}$ improves the temporal accuracy of the solution, at the expense of a linear increase in computational cost. We take throughout the text $n_{\rm steps}=640$. Third, at each timestep we solve a sparse linear system with the \texttt{SciPy} sparse solver \texttt{spsolve} \cite{Demmel:1999, SciPy:2020, Saad:2003, vanDerVorst:1992}.
\begin{figure*}[ht!]
  \centering
\includegraphics[width=0.49\linewidth]{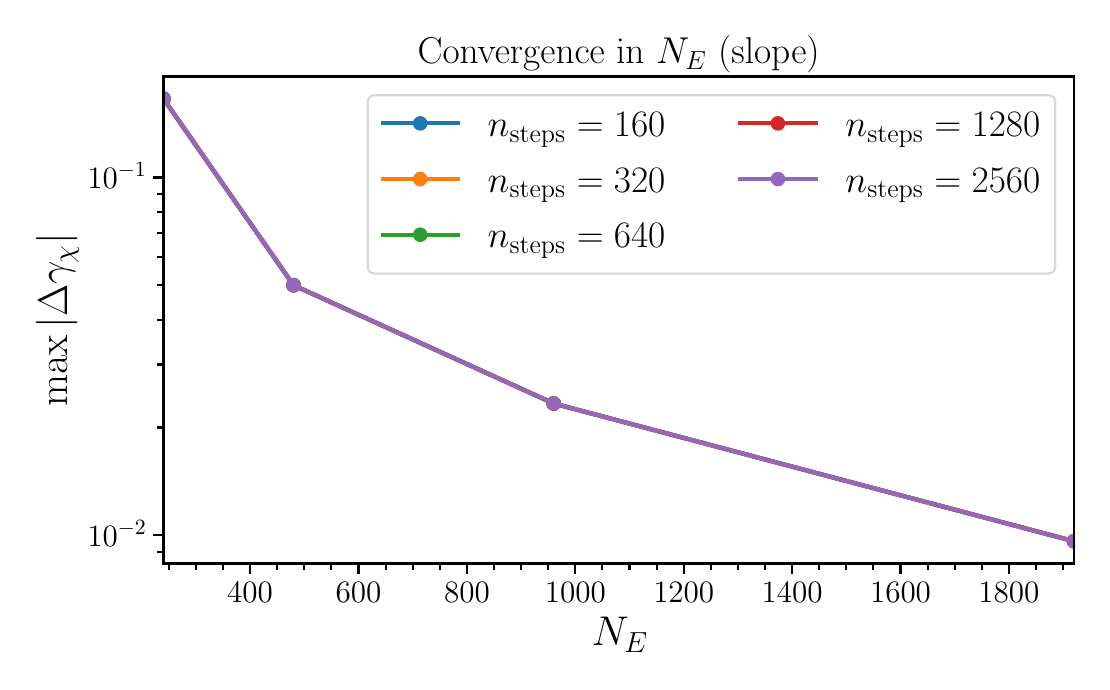} 
\includegraphics[width=0.49\linewidth]{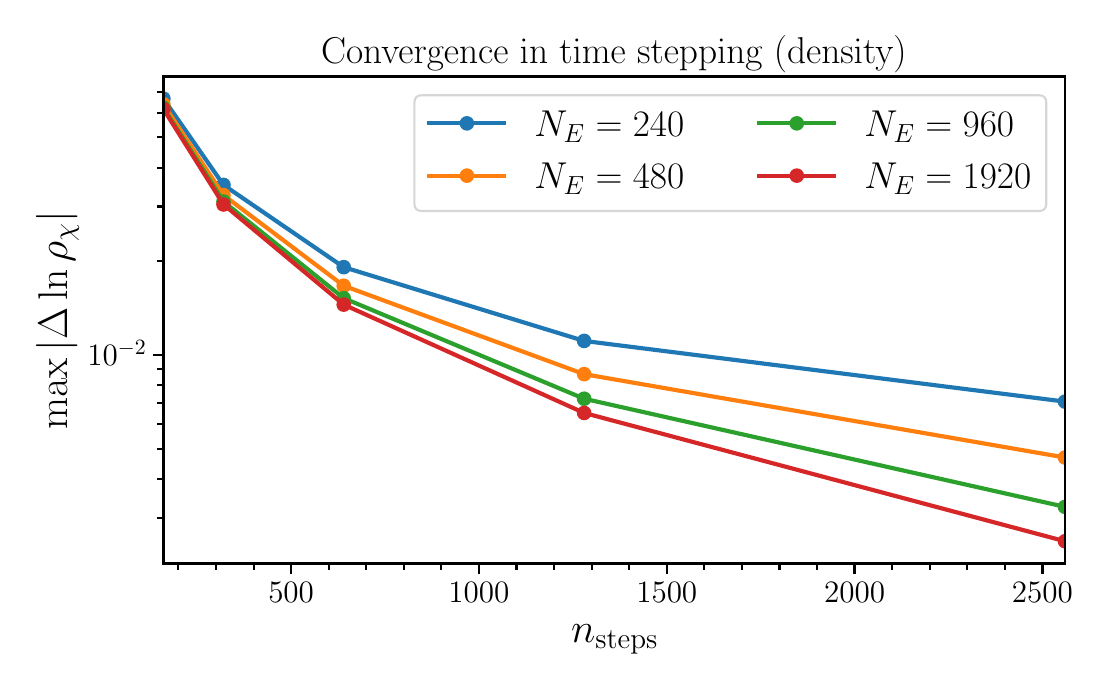} 
\includegraphics[width=0.49\linewidth]{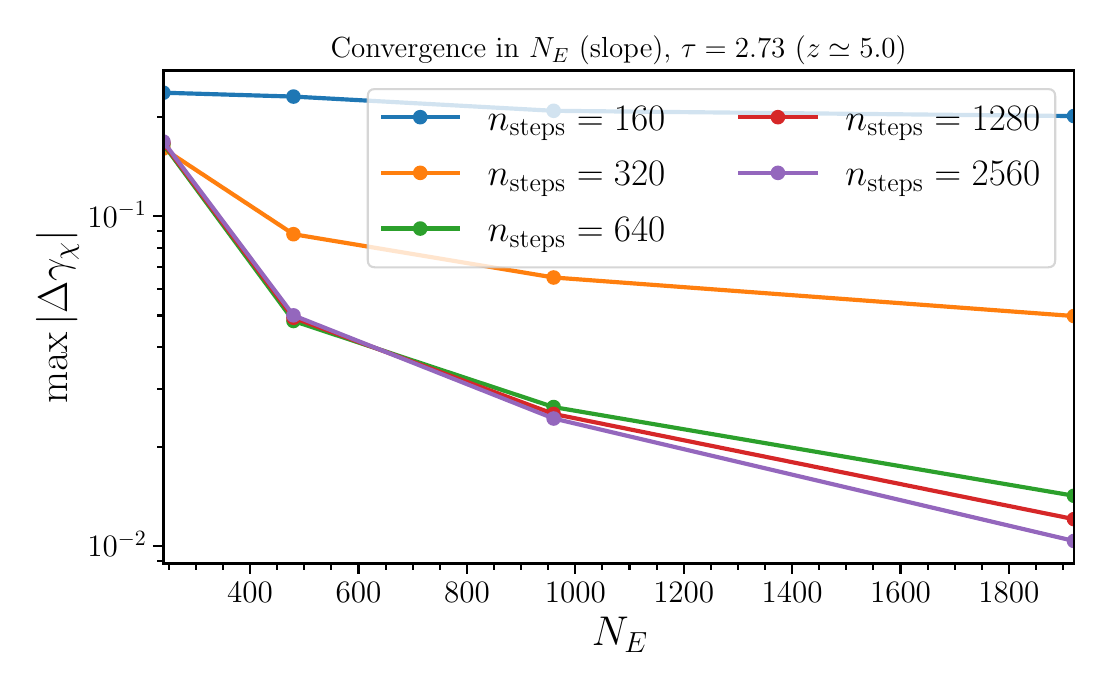} 
\includegraphics[width=0.49\linewidth]{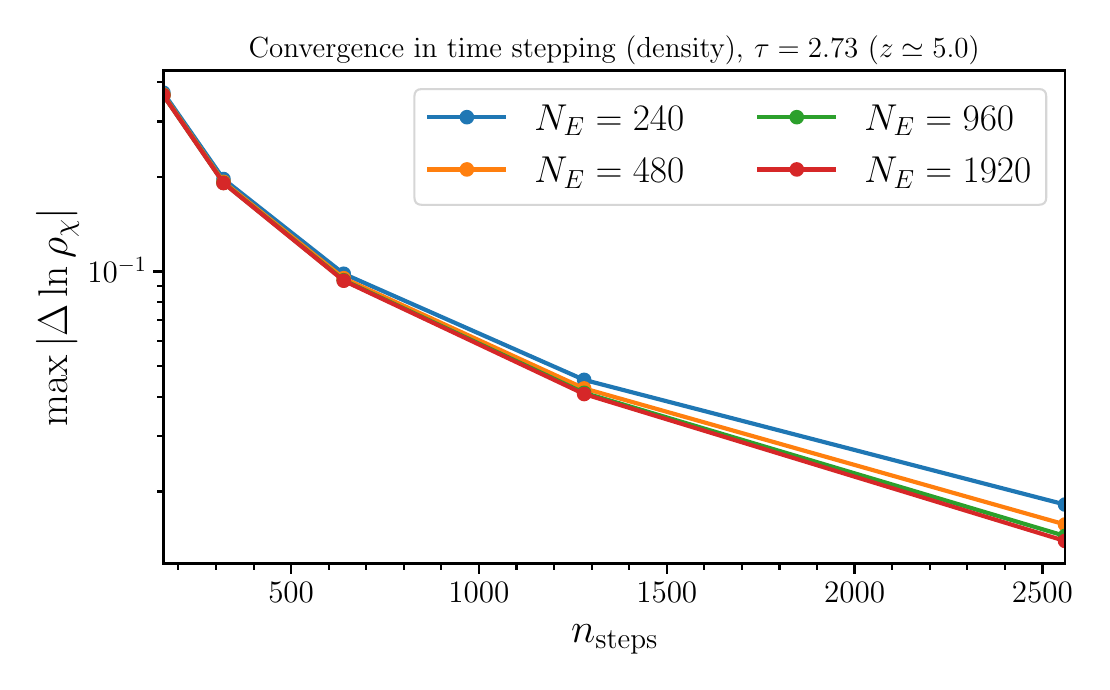} 
  \caption{\textit{Left panels}: Convergence of the DM density slope, quantified by the maximum absolute difference in the logarithmic slope $\gamma_\chi \equiv \mathrm{d}\ln\rho_\chi/\mathrm{d}\ln r$ within the radial range $r=10^{-3}$-$10^{-1}\,\mathrm{pc}$, as a function of the number of energy bins $N_E$, for several values of the number of timesteps $n_{\rm steps}$. Results are shown at $\tau=32$ ($z \simeq0.35$) in the upper panel, and at $\tau=2.73$ ($z \simeq5$) in the lower panel. The results at each point of the $N_E-n_{\rm steps}$ grid are compared to a high-resolution reference run with $N_E=2880$ and $n_{\rm steps}=5120$. \textit{Right panels}: Convergence of the DM density normalization, quantified by the maximum absolute difference in $\ln\rho_\chi$ within the same radial range, as a function of $n_{\rm steps}$ for different values of $N_E$.}
  \label{fig:convergence}
\end{figure*}
We asessed the impact of these numerical choices in two different parameters discussed throughout the work: the average density profile within the spike (for concreteness, we take the average over $r=10^{-3}-10^{-1}$pc) and the average density profile steepness within that same range. In this way, we tackle the impact of convergence parameters both on the normalization and steepness of the profile.

Regarding the tolerance parameter of the numerical solver, we find that this is much smaller than the uncertainty introduced by the choices of $N_E$ and $n_{\rm steps}$. We varied our SuperLU solver in python to an iterative Krylov solver, and also varied the convergence tolerance by orders of magnitude, finding that their relative impact on the average density and steepness is smaller than $\lesssim 10^{-5}$.

On the other hand, the choices of $N_E$ and $n_{\rm steps}$ do introduce a controlled but non-negligible numerical uncertainty. We quantify this effect for the coevolved stellar and DM Fokker-Planck system of Eq.~(\ref{eq:DE_star}) in Fig.~\ref{fig:convergence}. The left panel shows the maximum absolute difference in the logarithmic slope of the DM density profile, $\Delta\gamma_\chi \equiv \max|\gamma_\chi - \gamma_{\chi,\mathrm{ref}}|$, as a function of the number of energy bins $N_E$, for several values of $n_{\rm steps}$. The right panel shows the corresponding convergence in time stepping, quantified via the maximum difference in the logarithmic density, $\max|\Delta\ln\rho_\chi|$, as a function of $n_{\rm steps}$ for different values of $N_E$. For the reference results, we choose $N_{E,\rm ref}=2880$ and $n_{\rm steps,ref}=5120$.

We find that the numerical errors associated with energy-space discretization and time stepping are largely independent: once either $N_E$ or $n_{\rm steps}$ is sufficiently large, the remaining uncertainty is dominated by the other. For the choices adopted in the main text, $N_E=480$ and $n_{\rm steps}=640$, the DM density profiles slope is converged to better than $\mathcal{O}(10\%)$ throughout the radial range of interest. However, the uncertainty in the averaged density profiles is can be as large as a factor of $\sim 2$ in the radial region of interest (note we are plotting the maximal difference, therefore corresponding to the worst case scenario). Further increases mostly on $n_{\rm steps}$ lead to more precise results. For instance, taking $n_{\rm steps}=2560$ reduces the uncertainty in the average density down to $\sim 15\%$. Employing such precision would significantly slow down our calculations in the varied multi-dimensional parameter space shown in the main text, and therefore we restrict ourselves to $n_{\rm steps}=640$. We emphasize that these choices have a negligible impact on the density profile slopes, and thus do not change qualitatively (nor quantitatively, in a visually appreciable way in the Figures) the conclusions of this work.

\end{document}